\documentclass[12pt]{article}%
\usepackage{amsmath}
\usepackage{graphicx}
\usepackage{amsfonts}
\usepackage{amssymb}
\usepackage{setspace}
\providecommand{\U}[1]{\protect\rule{.1in}{.1in}}
\providecommand{\U}[1]{\protect \rule{.1in}{.1in}}
\providecommand{\U}[1]{\protect \rule{.1in}{.1in}}

\newtheorem{theorem}{Theorem}

\newtheorem{corollary}{Corollary}

\newtheorem{definition}{Definition}
\newtheorem{example}{Example}

\newtheorem{lemma}{Lemma}

\newtheorem{remark}{Remark}

\begin{document}

\title{Equilibria of nonatomic anonymous\ games\thanks{Acknowledgments to be added.}}
\author{Simone Cerreia-Vioglio$^{a}$, Fabio Maccheroni$^{a}$, David Schmeidler$^{b}$\\$^{a}${\small Universit\`{a} Bocconi and Igier, }$^{b}${\small Tel Aviv
University}}
\date{April\ 2020}
\maketitle

\begin{abstract}
We add here another layer to the literature on nonatomic anonymous games
started with the 1973 paper by Schmeidler. More specifically, we define a new
notion of equilibrium which we call $\varepsilon$-estimated equilibrium and
prove its existence for any positive $\varepsilon$. This notion encompasses
and brings to nonatomic games recent concepts of equilibrium such as
self-confirming, peer-confirming, and Berk--Nash. This augmented scope is our
main motivation. At the same time, our approach also resolves some conceptual
problems present in Schmeidler (1973), pointed out by Shapley. In that
paper\ the existence of pure-strategy Nash equilibria has been proved for any
nonatomic game with a continuum of players, endowed with an atomless countably
additive probability. But, requiring Borel measurability of strategy profiles
may impose some limitation on players' choices and introduce an exogenous
dependence among\ players' actions, which clashes with the nature of
noncooperative game theory. Our suggested solution is to consider every subset
of players as measurable. This leads to a nontrivial purely finitely additive
component which might prevent the existence of equilibria and requires a novel
mathematical approach to prove the existence of $\varepsilon$-equilibria.

\end{abstract}

\section{Introduction}

\subparagraph{The original framework of Schmeidler.}

Games with a continuum of anonymous players were introduced by Schmeidler in
\cite{Sch} where he also proved the existence of pure-strategy Nash equilibria
for these games.\footnote{Theorem 1 in \cite{Sch}\ is a special case of the
last theorem of Schmeidler's Ph.D. dissertation in mathematics titled
\textquotedblleft Games with a continuum of players\textquotedblright\ ;
submitted and approved in 1969, at the Hebrew University in Jerusalem. The
problem was inspired by the moonlighting job of the author as a member of a
team advising on Tel Aviv transportation.} At the time, there were models of
markets and cooperative games with infinitely many players, but not of
noncooperative games. In \cite{Sch}, the players' space is modelled to be the
unit interval endowed with the Borel $\sigma$-algebra and the Lebesgue
measure, where there is a finite set of actions and each player chooses an
action from this set. The utility of each player\ depends on the distribution
of actions\ across all players and the action he chooses. The interpretation
is that the same game is repeated in each period. The payoff, in utils, is
received at the end of the period. At the same time,\ because of the anonymity
assumption, the strategic complications of repeated games are meaningless
here. A paradigmatic example is that of daily commuters driving downtown (or
back home) and having\ to choose a bridge (or tunnel) to enter the city. Thus,
in each period they play a one-shot game, analyzed in\ \cite{Sch}. Here, the
metaphysical assumption of correctly guessing what other players will do,
required for playing a Nash\ equilibrium strategy in one-shot games, is
mitigated by two factors. The first is\ minor: each player has to guess
correctly the distribution of the strategy (the same guess for all). The
second is major: there is regularity in the daily traffic of commuters.
Schmeidler \cite{Sch} formalizes these\ intuitions. The limitations of this
model are discussed below.

\subparagraph{Our motivations.}

The goal of our paper is to generalize the above finding in several
directions. We are motivated by three main observations:

\begin{enumerate}
\item[(i)] In recent\ years, alternative, and perhaps more realistic, notions
of equilibrium have been developed for noncooperative games with finitely many
players. At the same time, these notions have not been considered for
nonatomic anonymous games. In particular, we have in mind equilibrium
concepts\ which allow for beliefs to be not necessarily correct, but
nonetheless consistent with the information possessed by each player
whether\ it is endogenously or exogenously generated. Thus, our goal is to
bring these more realistic notions of equilibrium to nonatomic anonymous games
which model exactly situations where individuals are negligible and are not
fully aware of the strategic environment surrounding them. This\ renders
sophisticated strategic reasoning, such as Nash equilibrium (and any of its
refinements) or rationalizability, less\ plausible.\footnote{We are not after
proving any sort of \textquotedblleft translation principle\textquotedblright,
that is, a principle for which any equilibrium notion developed for a
finite-players framework easily translates, in terms of existence, to\ a
nonatomic setting.}

\item[(ii)] In a personal conversation\ with Schmeidler (in the early 1970s),
Shapley pointed out a problem with the modelling of a
nonatomic\ population\ of players as the unit interval with the Lebesgue
measure on Borel sets. As in some mathematical sense there are more
nonmeasurable sets than measurable sets in the unit interval, the game, that
is the payoff function, may not be defined out of equilibrium. In a similar
vein, as later formalized in a general equilibrium framework, Dubey and
Shapley \cite{DS} raise another issue with the measurability assumption. The
measurability of a strategy profile (and similarly of the profile of
utilities, which is a common assumption) yields its \textquotedblleft
near\textquotedblright\ continuity.\footnote{This intuition is based on
Lusin's Theorem which states that for each $\varepsilon>0$ each measurable
function is continuous when restricted to a suitable compact set which\ has a
measure of at least\ $1-\varepsilon$ (see, e.g., Aliprantis and Border
\cite[Theorem 12.8]{AB}).} This in turn clashes with\ the noncooperative idea
of strictly independent decision-making, since \textquotedblleft close
players\textquotedblright\ tend to play \textquotedblleft close
strategies\textquotedblright.

\item[(iii)] In modelling a large population of players in which each agent
\textquotedblleft has the same negligible weight\textquotedblright, Schmeidler
opted for the infinite set of points in the unit interval endowed with the
Lebesgue measure. At the same time, as noted by Aumann \cite{Aum}, in
analyzing economies with a continuum of traders, \textquotedblleft the choice
of the unit interval as a model for the set of [players] is of no particular
significance. A planar or spatial region would have done just as well. In
technical terms, [the players' space] can be any measure space without
atoms.\textquotedblright\ Thus, for example, one could alternatively\ model
the players' space as the set of natural numbers endowed with a natural
density. Our goal is to take Aumann's remark\ verbatim and not commit to any
particular specification of the players' space in order to see how much of our
analysis can be carried out in a general space without atoms. More formally,
we suggest using\ Savage's structure of nonatomic probabilities defined on the
power set of the space of players (Section \ref{sec:pla-spa}).
\end{enumerate}

\subparagraph{Our contributions.}

Our second and third motivation bring us to model the\ players' space as a set
$T$ endowed with a nonatomic probability $\lambda$ defined on all subsets $T$.
Using a measure over the power set takes care of both Shapley's and Aumann's
comments. In particular, by considering the power set, we allow for the most
permissive measurable structure possible, since any profile of strategies or
utilities becomes automatically measurable.\ Measuring the subsets of
players/coalitions according to a nonatomic probability on the power set is
consistent with Savage's\ \cite{Sav} approach and equivalent to having a
qualitative probability on the players' space, satisfying axiom P6'. However,
modelling the players'\ space in this generality implies that Nash equilibria
might fail to exist (see Example \ref{exa:KPQS} based on Khan, Qiao, Rath, and
Sun \cite{KQRS}).

This naturally brings us to look at $\varepsilon$-equilibria and to our first
motivation. We introduce a concept of approximate equilibrium for nonatomic
anonymous games, which we call $\varepsilon$-estimated equilibrium. This
notion of $\varepsilon$-equilibrium encompasses several approximate
equilibrium concepts: $\varepsilon$-self-confirming ($\varepsilon$-SCE),
$\varepsilon$-peer-confirming ($\varepsilon$-PCE), and $\varepsilon
$-Berk--Nash ($\varepsilon$-BNE). These equilibria\ and their $\varepsilon
$-versions are formally defined and discussed in the relevant sections,
Sections \ref{sec:msg-gam}, \ref{sec:nei-gam}, and \ref{sec:mis-gam} (see also
the related literature below). They were\ mostly developed for finite games
and, inter alia, in this paper we extend them to nonatomic games.
Nevertheless, the principles behind their\ definitions in a finite-players
framework naturally translate to a nonatomic setup. The common thread behind
$\varepsilon$-SCE, $\varepsilon$-PCE, and $\varepsilon$-BNE in an anonymous
nonatomic game is the following scheme, which is also the basis for our
$\varepsilon$-estimated equilibria:

\begin{enumerate}
\item Every player best-responds to his beliefs (optimality);

\item The belief of every player is consistent\ with what he can\ observe
($\varepsilon$-discrepancy).\footnote{More precisely, we require points 1 and
2 to hold for every player except a null set of them (see also point 1 of
Remark \ref{rmk:RSCE}).}
\end{enumerate}

Where these types of equilibrium differ is how point 2 is formalized, since
point 1 is translated in the same way for all of them. In particular, in SCE,
each player receives a message which is a function of the action he takes and
the distribution of actions of the other players. In equilibrium, almost every
player best-responds to a distribution that generates a message which is
$\varepsilon$-close to the message generated by the true distribution of the
actions. In PCE, the message each player receives is the distribution of the
actions conditional on a subset of players: his peers. Thus, almost all the
players best-respond to a distribution which is $\varepsilon$-close to the
true distribution of actions of their peers, not of all the players. In both
$\varepsilon$-SCE and $\varepsilon$-PCE the distributions to which players
best-respond are $\varepsilon$-close in terms of observables to the true one;
thus they are \emph{endogenously} generated. By contrast, in BNE, each player
$t$ entertains an \emph{exogenous} set of possible distributions of actions,
denoted by $Q_{t}$, that he believes are accurate in describing other players'
behavior. Moreover, he is not willing to depart from $Q_{t}$. So in
equilibrium, almost every player best-responds to a distribution which is
$\varepsilon$-close to the best estimate in $Q_{t}$ of the true distribution
of actions, according to a statistical measure.

Our notion of $\varepsilon$-estimated equilibrium provides a framework where
we can account for all the three different features described above: that is,
the distribution of actions used by each player in equilibrium is
$\varepsilon$-close, whether in statistical terms or proper distance, to the
set of all distributions which are compatible with the true one. This latter
set can be exogenously determined as in BNE or endogenously generated as in
SCE or PCE.

In Theorem \ref{thm:mai}, under mild assumptions, we prove that $\varepsilon
$-estimated equilibria always exist. As particular cases, we obtain the
existence of self-confirming $\varepsilon$-equilibria (Corollary
\ref{cor:eps-sce}), peer-confirming $\varepsilon$-equilibria (Corollary
\ref{cor:eps-pce}), and Berk--Nash $\varepsilon$-equilibria\ (Corollary
\ref{cor:eps-bne}). Despite the fact that standard Nash equilibria might fail
to exist, we prove that $\varepsilon$-Nash equilibria do exist (Corollary
\ref{cor:eps-Nas}). Finally, mimicking the notion of rationalizable
self-confirming equilibrium\ (see Rubinstein and Wolinsky \cite{RW}), we also
propose a definition of rationalizable estimated equilibrium and discuss its
existence (Remark \ref{rmk:RSCE}).

\subparagraph{Related literature.}

The seminal contribution of Aumann \cite{Aum}\ (in a general equilibrium
framework), followed by Schmeidler \cite{Sch} (in a game-theoretic framework),
initiated a large literature where the negligibility of agents is modelled
via\ a nonatomic probability players' space (see, e.g., Khan and Sun
\cite{KS2}\ for a survey).\footnote{Many subsequent papers extended
Schmeidler's\ results to more general players' spaces, but where $\lambda$ is
always assumed to be countably additive and $A$ is allowed to be infinite:
see, e.g., Balder \cite{Bal}, Khan and Sun \cite{KS}, Khan, Rath, and Sun
\cite{KRS}, Rath \cite{Rat}, and the references therein. The scope of
this\ type of results is analyzed in Carmona and Podczeck \cite{CP}. Finally,
in the same setting of Schmeidler \cite{Sch}, Jara-Moroni \cite{Jar} extends
the notion of rationalizability to nonatomic anonymous games while Rath
\cite{Rat1} investigates the issue of existence of perfect, proper, and
persistent equilibria, being all of them refinements of Nash equilibrium.} We
will next discuss the relevant literature by connecting it to our three main motivations/contributions.

\begin{itemize}
\item[(i)] Our definition\ of $\varepsilon$-estimated equilibrium\ seems to be
new. At the same time, it encompasses three types\ of equilibrium:
self-confirming (SCE), peer-confirming (PCE), and Berk--Nash (BNE) which were
developed almost exclusively for games with finitely many players,
respectively, by Battigalli \cite{Bat} as well as Fudenberg and Levine
\cite{FL} (SCE), Lipnowski and Sadler \cite{LS} (PCE), and Esponda and Pouzo
\cite{EP} (BNE). The only exceptions seem to be SCE and BNE, which were also
studied\ for population games, where the latter can be seen as a very special
form of nonatomic games. Moreover, we also consider $\varepsilon$-versions of
the above three concepts of equilibrium. In discussing $\varepsilon$-SCE of
course, two approaches are available. The first assumes that: a) players
best-respond to their beliefs, but b) beliefs are only $\varepsilon
$-consistent with evidence. The second requires that: a') players
$\varepsilon$-best-respond to their beliefs, but b') beliefs are
perfectly\ consistent. For games with finitely many players, the first
approach was introduced by Battigalli \cite{Bat} and Kalai and Lehrer
\cite{KL} and \cite{KL1}, while the second was proposed for pure equilibria by
Azrieli \cite{Azr}. For nonatomic games, other than population games, the
first approach seems to be unexplored, while the second was studied by Azrieli
\cite{Azr}. Using the same setting as Schmeidler, that is, assuming that the
players' space is the unit interval with the Lebesgue measure, Azrieli shows
that self-confirming equilibria exist (that is, when $\varepsilon=0$), but
when utility depends on the entire profile of strategies and the message
feedback is the distribution of actions.\footnote{In our specification, this
would collaps to a Nash equilibrium.} Moreover, in trying to obtain the
nonatomic games of Schmeidler as a limit of finite-players games which become
arbitrarily large, he shows that self-confirming $\varepsilon$-equilibria
eventually exist.\footnote{For a related concept and result see also Section 5
of Fudenberg and Kamada \cite{FK}.} Finally, Azrieli limits his analysis to
the case where there is nonmanipulable information also known as own-action
independence of feedback. Loosely speaking, this is the case when the feedback
each player receives does not depend on the action taken by the player. This
rules out several interesting cases.

In our work, we opt for a definition\ of $\varepsilon$-SCE which requires
rational optimization on\ the players' side, but allows them to entertain
$\varepsilon$-consistent\ beliefs. We do not assume own-action independence.
The assumption of\ $\varepsilon$ being strictly positive is due to two
reasons: one mathematical and one conceptual. Mathematically, by considering
players' spaces which involve finitely additive measures $\lambda$, one can
show that self-confirming equilibria might fail to exist (Example
\ref{exa:KPQS}). Conceptually, we take the point of view of Kalai and Lehrer
\cite{KL} and \cite{KL1}: we impose rational behavior on players, but allow
for slightly inconsistent beliefs. The latter assumption can be justified by
interpreting the belief of each player as the belief entertained after many
rounds of play, so that learning yields approximately correct predictions
about observables. At the limit, beliefs would be perfectly consistent with
observations, but before that they might be just $\varepsilon$-consistent.

\item[(ii)] The issue of measurability in nonatomic economies and games has
been raised and dealt with by several authors in the past. Khan and Sun
\cite{KS1} proposed to replace the unit interval with the Lebesgue
measure\ with a generic Loeb space. In this way, players (resp., coalitions)
are represented as hyperreals (resp., sets of hyperreals). Their approach is
mathematically very elegant, but very different from ours. Ours is
conceptually simpler: we simply remove any measurability constraint by
replacing the Borel $\sigma$-algebra\ with the power set. This comes at a
cost:\ the loss of countable additivity of $\lambda$. This not only
complicates the technical analysis, but generates a conceptual loss. In fact,
in an independent paper, Khan, Qiao, Rath, and Sun \cite{KQRS} show that the
existence of Nash equilibria for any game with players' space $\left(
T,\mathcal{T},\lambda\right)  $ is equivalent to the countable additivity of
$\lambda$. Since the existence of Nash equilibria cannot be guaranteed with
mere finite additivity, they study the existence of $\varepsilon$-Nash
equilibria, thus overlapping our Corollary \ref{cor:eps-Nas}.

\item[(iii)] The issue of modelling the players' space as a continuum or as a
discrete space has also been discussed by Al-Najjar \cite{AlN}, who\ considers
as competing models the continuum space $\left[  0,1\right]  $ versus\ a dense
countable grid of $\left[  0,1\right]  $. This paper also shares some of the
motivation coming from the Dubey-Shapley's remark on measurability (see Dubey
and Shapley \cite{DS} as well as Khan and Sun \cite{KS1}). Thus, in trying to
build a link between these two conceptually equivalent approaches countable
additivity is necessarily lost, as in our case. The main results of
Al-Najjar\ \cite{AlN} show that, under suitable conditions, the two approaches
to\ modelling the players' space, that is, a continuum versus\ a discrete
dense grid,\ are equivalent. In order to achieve this result, Al-Najjar shows
that all his Nash-type equilibria, for his class of discrete games, can be
purified. Compared to our work, Al-Najjar is not concerned with any other
form\ of equilibrium other than Nash equilibria. Moreover, he establishes the
existence of a form of $\varepsilon$-equilibrium for those discrete nonatomic
games that arise as limits of proper sequences of finite-players games.
Example \ref{exa:AlN} shows that for our more general class of games these
$\varepsilon$-equilibria are not always guaranteed to exist.
\end{itemize}

We conclude by mentioning one more work. One of the important papers on
nonatomic games which introduces a novel approach is Mas-Colell \cite{Mas}.
His approach is based on distributions of strategies,\footnote{This
reformulation is connected to the distributional approach for Bayesian games
with a continuum of types (see \cite[Remarks 3 and 4]{Mas}).} which allows for
not considering strategy profiles. In this way, issues\ of measurability can
be partially overridden\ in the proofs. It is an alternative framework which
permits the discussion of players' negligibility. In this\ framework though,
Shapley's observation would\ still apply and the assumption of countable
additivity still seems to be playing a major role. Finally, we are not aware
of refinements of and variations on this distributional concept of equilibrium.

\subparagraph{Roadmap.}

In Sections \ref{sec:pla-spa} and \ref{sec:sta-gam}, we formally introduce
nonatomic players' spaces, nonatomic\ games with estimation feedback, and the
definition\ of $\varepsilon$-estimated equilibrium whose existence is proven
in Theorem \ref{thm:mai}. In Sections \ref{sec:msg-gam}, \ref{sec:nei-gam},
and \ref{sec:mis-gam}, as a by-product, we obtain the existence of
self-confirming, Nash,\ peer-confirming, and Berk--Nash $\varepsilon
$-equilibria. Proofs are relegated to the appendices. In particular, in
Appendix \ref{app:bel}, Lemma \ref{lem:believe-me} generalizes Theorem 7 of
Khan and Sun \cite{KS}\ which deals with the set of distributions induced by
all the selections of a correspondence. In Appendix \ref{app:pro}, we provide
a brief summary of\ how the main proofs are carried out and prove all the
results contained in the main text.

\section{Nonatomic players' spaces\label{sec:pla-spa}}

A\ \emph{players' space} is a pair $\left(  T,\lambda\right)  $ where $T$\ is
a set of players and $\lambda$ is a (finitely additive) \emph{probability\ }on
the power set of $T$.\footnote{Recall that $\lambda$ is a finitely additive
probability if and only if $\lambda$ is a positive finitely additive set
function such that $\lambda\left(  T\right)  =1$.}\ When $T=\mathbb{N}$, a
fundamental class\ of probabilities that are not countably additive are
\emph{natural densities}, that is, probabilities $\lambda$ such that%
\[
\lambda\left(  E\right)  =\lim_{k\rightarrow\infty}\frac{\left\vert
E\cap\left\{  1,...,k\right\}  \right\vert }{k}%
\]
whenever the limit exists. As is well known, there are many natural densities
and all of them satisfy the following property:

\smallskip

\noindent\textbf{Strong continuity (Savage's nonatomicity)}\emph{\ For each
}$\varepsilon>0$\emph{\ there exists a finite partition }$\left\{  F_{1}%
,F_{2},...,F_{k}\right\}  $\emph{\ of }$T$\emph{\ such that }$\lambda\left(
F_{i}\right)  <\varepsilon$\emph{\ for all }$i=1,...,k$\emph{.}

\smallskip

Under strong continuity, any singleton (i.e., any single player) has measure
$0$ and for each $F\subseteq T$\ and $\beta\in\left(  0,1\right)  $ there
exists $E\subseteq F$ such that $\lambda\left(  E\right)  =\beta\lambda\left(
F\right)  $.\footnote{See Maharam \cite[Example 2.1 and Theorem 2]{Mah} and
Bhaskara Rao and Bhaskara Rao \cite[Theorem 5.1.6 and Remark 5.1.7]{RR}. In
this literature, natural densities are called density measures or
density\ charges.} This is the class of probabilities introduced by
Savage\ \cite{Sav} when he solved De Finetti's open problem on the
representation of qualitative probabilities (see also Samet and Schmeidler
\cite{SaS}).

\section{Nonatomic games and their equilibria\label{sec:sta-gam}}

Nonatomic games are games where each single player has no influence on the
strategic interaction, but only the aggregate\ behavior of \textquotedblleft
large\textquotedblright\ sets of players can change the players' payoffs.
Formally, a \emph{nonatomic (anonymous) game }is a triplet\ $G=\left(  \left(
T,\lambda\right)  ,A,u\right)  $ where $\left(  T,\lambda\right)  $ is the
players'\ space, $A$ is the space of players' actions/strategies and $u\ $is
their profile of utilities.\footnote{In the paper, given a generic set $B$, we
use the term profile to refer to a function from the set of players $T$ to
$B$. We will denote a profile by either $b:T\rightarrow B$ or by $b=\left(
b_{t}\right)  _{t\in T}$. The latter notation will allow us, with a
small\ abuse, to treat $\left(  b_{t}\right)  _{t\in T}$ also as a set.}
Below, we discuss in detail these mathematical objects and their interpretations.

\begin{itemize}
\item $A=\left\{  1,...,n\right\}  $ is\ the\ set of pure
\emph{strategies/actions}.

\item $\Delta=\left\{  x\in\mathbb{R}_{+}^{n}\mid\sum_{i=1}^{n}x_{i}%
=1\right\}  $ is the $n-1$ dimensional simplex. We denote by $d_{\Delta}$ the
distance on $\Delta$ induced by the Euclidean norm. This set\ represents all
possible \emph{distributions of players' strategies}. Note that an element in
$\Delta$ can actually take two possible interpretations. In fact, given a
player $t$ and an element of $\Delta$, this element\ can either be interpreted
as a subjective belief of player $t$ (in this case, we often denote it by
$\beta_{t}$) or be interpreted as an\ objective distribution of players'
strategies (in this case, we typically denote it by $x$).

\item $u=\left(  u_{t}\right)  _{t\in T}$ is a profile\ of functions
$u_{t}:A\times\Delta\rightarrow%
\mathbb{R}
$. For each $t$ in $T$, $u_{t}\left(  a,\beta_{t}\right)  $ represents the
ex-ante\ \emph{utility} of player $t$, when he chooses strategy $a$, if his
belief about the distribution of opponents' strategies is $\beta_{t}$.
\end{itemize}

As mentioned in the Introduction, nonatomic games were first studied by
Schmeidler \cite{Sch}. In this paper, we consider a class of games which we
term \emph{nonatomic\ games with estimation feedback}. It has a richer
structure and nonatomic games can be seen as a specific parametrization.

Formally, a nonatomic game with estimation feedback is a
quintet\ $G=((T,\lambda),A,u,(\Pi,\pi),f)$\ where $\left(  \left(
T,\lambda\right)  ,A,u\right)  $ is a nonatomic\ game defined as above,
$\left(  \Pi,\pi\right)  $ is a neighborhood structure, and $f$ is a profile
of estimation feedback functions which discipline the beliefs' formation of
agents in equilibrium. Formally, we have that:

\begin{itemize}
\item $\left(  \Pi,\pi\right)  $ is a\ \emph{neighborhood structure} if and
only if\ $\Pi=\left\{  T_{j}\right\}  _{j=1}^{m}$ is a finite cover of $T$
whose elements have strictly positive measure and $\pi$ is a function from $T$
to $\left\{  1,...,m\right\}  $. In particular, each $T_{j}$ is a nonempty
subset of $T$ such that\ $\lambda\left(  T_{j}\right)  >0$ and $\cup_{j=1}%
^{m}T_{j}=T$. An important example of finite covers are finite partitions of
the players' space. We interpret an element of $\Pi$, $T_{j}$, as the $j$-th
subpopulation of $T$ and for each $t\in T$ the value $\pi\left(  t\right)  $
will denote which subpopulation player $t$ observes.\footnote{Despite being a
natural requirement, we can dispense with the assumption that $t\in
T_{\pi\left(  t\right)  }$. In other words, we do not need to assume that any
player $t$ belongs to the subpopulation he observes.}

\item $f=\left(  f_{t}\right)  _{t\in T}$ is a profile of \emph{(estimation)}
\emph{feedback functions}\ $f_{t}:A\times\Delta\times\Delta\rightarrow\left[
0,\infty\right)  $. Each $f_{t}$ is assumed to be such that for each
$y\in\Delta$ there exists $x_{y}\in\Delta$\ for which it holds that%
\begin{equation}
f_{t}\left(  a,x_{y},y\right)  =0\qquad\forall a\in A \label{eq:wea-gro}%
\end{equation}
For each $t$ in $T$, $f_{t}\left(  a,\beta_{t},x\right)  $\ represents a
measure of consistency between the belief $\beta_{t}$ (entertained by player
$t$) about the players' actions within\ the subpopulation observed by $t$\ and
the actual distribution of players' strategies $x$ within\ that same
subpopulation, with the idea that the larger $f_{t}\left(  a,\beta
_{t},x\right)  $ is the greater\ is the discrepancy between the player's
belief and the subpopulation actions' distribution.\ In line with this
interpretation, property (\ref{eq:wea-gro}) says that for each possible true
model $x$ there exists a belief $\beta_{t}$ such that this discrepancy is
minimal, no matter what action $a$\ is chosen by player $t$. To better
understand (\ref{eq:wea-gro}), we next state a stronger property which implies
(\ref{eq:wea-gro})\ and has a more immediate interpretation. In all our
specifications, with the exception of (\ref{eq:pro-div}), it will be
satisfied: for each $t\in T$ and for each $a\in A$%
\begin{equation}
x=y\implies f_{t}\left(  a,x,y\right)  =0 \label{eq:gro}%
\end{equation}
In words, this latter property says that discrepancy is minimal provided the
belief $\beta_{t}$ is indeed\ correct, that is $\beta_{t}=x$.\footnote{Note
that (\ref{eq:gro}) implies (\ref{eq:wea-gro}). Fix $t\in T$.\ For each
$y\in\Delta$, set $x_{y}=y$. By (\ref{eq:gro}), it follows that $f_{t}\left(
a,x_{y},y\right)  =0$ for all $a\in A$.}\ Under (\ref{eq:wea-gro})
or\ (\ref{eq:gro}), we deliberately allow for the possibility that
$f_{t}\left(  a,\beta_{t},x\right)  =0$, but $\beta_{t}\not =x$:\ a belief
might be consistent with evidence, but still incorrect.
\end{itemize}

Finally, we need three extra mathematical objects:

\begin{itemize}
\item $\Sigma=A^{T}$ is the set of all functions $\sigma$ from $T$ to $A$.
Each $\sigma\in\Sigma$ represents a \emph{strategy profile} in which the
generic player $t$ chooses strategy\ $\sigma\left(  t\right)  $.

\item Given $j\in\left\{  1,...,m\right\}  $, $\lambda^{j}$ denotes the
probability on the power set of $T$ defined by%
\[
\lambda^{j}\left(  E\right)  =\frac{\lambda\left(  E\cap T_{j}\right)
}{\lambda\left(  T_{j}\right)  }\qquad\forall E\subseteq T
\]
In other words, $\lambda^{j}$ is the players' conditional measure\ in\ the
subpopulation $j$. Note that if $\lambda$ is strongly continuous, so is each
$\lambda^{j}$.

\item Given $\sigma\in\Sigma$ and $j\in\left\{  1,...,m\right\}  $,
$\lambda_{\sigma}^{j}\in\Delta$ is the distribution of $\sigma$ on $A$ in the
$j$-th subpopulation,\ that is,%
\[
\lambda_{\sigma}^{j}=\left(  \lambda^{j}\left(  \left\{  t\in T_{j}\mid
\sigma(t)=a\right\}  \right)  \right)  _{a\in A}%
\]
The vector $\lambda_{\sigma}^{j}$ represents the \emph{true} distribution of
players' pure strategies in\ the $j$-th subpopulation,\footnote{By definition
of $\lambda_{\sigma}^{j}$, note that
\[
\lambda_{\sigma}^{j}=\left(  \lambda^{j}\left(  \left\{  t\in T\mid
\sigma(t)=a\right\}  \right)  \right)  _{a\in A}%
\]
for all $j\in\left\{  1,...,m\right\}  $.} when they all play according to
$\sigma$. When $\Pi$ is trivial, that is, $\Pi=\left\{  T\right\}  $, then
$\Pi$ contains only one element and $\lambda=\lambda^{1}$. In this case, we
write $\lambda_{\sigma}$ in place of $\lambda_{\sigma}^{1}$. Similarly,\ the
vector $\lambda_{\sigma}$ represents the true distribution of players' pure
strategies in the entire population.
\end{itemize}

\bigskip

We can now introduce our most general concept of equilibrium. It provides a
unifying structure for the notions of equilibrium that feature players best
responding to beliefs that are possibly wrong, but are nonetheless consistent
with\ their\ probabilistic information. In the next three sections, we discuss
three\ particular and important\ specifications (see also the Introduction).

\begin{definition}
\label{def:eps-sta}Let $\varepsilon\geq0$. \emph{An\ }$\varepsilon
$\emph{-estimated equilibrium (in pure strategies) }for the nonatomic game
with estimation feedback $G=\left(  \left(  T,\lambda\right)  ,A,u,\left(
\Pi,\pi\right)  ,f\right)  $ is a strategy profile $\sigma\in\Sigma$ such that
there exists a profile of beliefs $\beta\in\Delta^{T}$ satisfying%
\begin{equation}
\lambda\left(  \left\{  t\in T\ \left\vert \
\begin{array}
[c]{l}%
u_{t}\left(  \sigma\left(  t\right)  ,\beta\left(  t\right)  \right)  \geq
u_{t}\left(  a,\beta\left(  t\right)  \right)  \quad\forall a\in A\\
f_{t}\left(  \sigma\left(  t\right)  ,\beta\left(  t\right)  ,\lambda_{\sigma
}^{\pi\left(  t\right)  }\right)  \leq\varepsilon
\end{array}
\right.  \right\}  \right)  =1 \label{eq:e-sta}%
\end{equation}

\end{definition}

We are ready to state our main result.

\begin{theorem}
\label{thm:mai}Let $G=\left(  \left(  T,\lambda\right)  ,A,u,\left(  \Pi
,\pi\right)  ,f\right)  $ be a nonatomic game with estimation feedback and
$\varepsilon>0$.\ If $\lambda$ is strongly continuous\ and $f=\left(
f_{t}\right)  _{t\in T}$ is a family of functions which is equicontinuous with
respect to the third argument,\footnote{We say that $f=\left(  f_{t}\right)
_{t\in T}$ is a family of functions which is equicontinuous with respect to
the third argument if and only if for each $\varepsilon>0$ there exists
$\delta_{\varepsilon}>0$ such that%
\[
d_{\Delta}\left(  x,y\right)  <\delta_{\varepsilon}\implies\left\vert
f_{t}\left(  a,\gamma,x\right)  -f_{t}\left(  a,\gamma,y\right)  \right\vert
<\varepsilon\quad\forall t\in T,\forall a\in A,\forall\gamma\in\Delta
\]
In other words, the family of functions $\left\{  f_{t}\left(  a,\gamma
,\cdot\right)  \right\}  _{t\in T,a\in A,\gamma\in\Delta}$\ from $\Delta$ to
$\left[  0,\infty\right)  $ is equicontinuous.}\ then $G$ has an $\varepsilon
$-estimated equilibrium.
\end{theorem}

\begin{remark}
\label{rmk:RSCE}\emph{Three observations are in order:}

\begin{enumerate}
\item \emph{In proving Theorem \ref{thm:mai}, we actually show that there
exists an }$\varepsilon$\emph{-estimated equilibrium} \emph{in which} each
\emph{player best-responds to his }$\varepsilon$\emph{-discrepant belief (cf.
also Remark \ref{ann:oso-sta}\ and Lemma \ref{lem:BR-equ-sta}), that is,\ the
set\ in (\ref{eq:e-sta}) coincides with\ }$T$\emph{ and, in particular, has
measure }$1$\emph{.}

\item \emph{As just mentioned, in an }$\varepsilon$\emph{-estimated
equilibrium players best-respond to their }$\varepsilon$\emph{-discrepant
beliefs. Mimicking the notion of\ rationalizable self-confirming
equilibrium\ of Rubinstein and Wolinsky \cite{RW}, we could also require that
this is correctly and commonly believed by all players. This will turn out to
be useful in discussing peer-confirming equilibrium (see Section
\ref{sec:nei-gam}).\ In order to do so, we first introduce some notation and
then propose a recursive definition. Given a nonempty subset }$S\subseteq
\Sigma$\emph{, we denote by }$\Delta\left(  S\right)  $ \emph{the set of all
probabilities over the power set of }$S$\emph{. Consider a player }$t\in
T$\emph{. An element }$\tilde{\beta}_{t}\in\Delta\left(  S\right)  $\emph{
represents the belief of the player about which strategy profile in\ }%
$S$\emph{ will realize.\ At the same time, given our assumption of anonymity
and the neighborhood structure, what is relevant for }$t$\emph{\ is merely the
distribution of players' strategies }$\bar{\beta}_{t}$, \emph{induced by
}$\tilde{\beta}_{t}$, \emph{within the subpopulation
observed.\footnote{Formally, we have that%
\[
\bar{\beta}_{t}=\left(  \int_{S}\lambda^{\pi\left(  t\right)  }\left(
\left\{  t\in T_{\pi\left(  t\right)  }\mid\sigma(t)=a\right\}  \right)
d\tilde{\beta}_{t}\right)  _{a\in A}%
\]
or, more succinctly,\ $\bar{\beta}_{t}=\int_{S}\lambda_{\sigma}^{\pi\left(
t\right)  }d\tilde{\beta}_{t}$.}\ With this, given }$\delta,\varepsilon\geq
0$\emph{, we can define recursively the following sequence\ of sets }$\left\{
S_{k}\right\}  _{k\in%
\mathbb{N}
_{0}}$\emph{: }$S_{0}=\Sigma$\emph{ and}%
\[
S_{k+1}=\left\{  \sigma\in S_{k}\mid%
\hspace{-1mm}%
\exists\tilde{\beta}\in\Delta\left(  S_{k}\right)  ^{T}\text{\emph{ s.t.\ }%
}\forall t\in T%
\begin{array}
[c]{c}%
u_{t}\left(  \sigma\left(  t\right)  ,\bar{\beta}\left(  t\right)  \right)
\geq u_{t}\left(  a,\bar{\beta}\left(  t\right)  \right)  -\delta\ \forall
a\in A\\
f_{t}\left(  \sigma\left(  t\right)  ,\bar{\beta}\left(  t\right)
,\lambda_{\sigma}^{\pi\left(  t\right)  }\right)  \leq\varepsilon
\end{array}
\right\}
\]
\emph{We say that }$\sigma\in\Sigma$ \emph{is a rationalizable\ }$\left(
\delta,\varepsilon\right)  $-estimated equilibrium (in pure strategies)\emph{
for the nonatomic game with estimation feedback }$G=\left(  \left(
T,\lambda\right)  ,A,u,\left(  \Pi,\pi\right)  ,f\right)  $\emph{ if and only
if\ }$\sigma\in\cap_{k\in%
\mathbb{N}
_{0}}S_{k}$\emph{. In words, in a rationalizable }$\left(  \delta
,\varepsilon\right)  $\emph{-estimated equilibrium, players }$\delta
$\emph{-best-respond to their }$\varepsilon$\emph{-discrepant beliefs and this
is correctly and commonly believed by all players. By setting }$\varepsilon
=\delta=0$\emph{, our definition reduces\ to a version\ for nonatomic
anonymous games of the equilibrium notion of Rubinstein and Wolinsky
\cite{RW}. We discuss existence in the next point.}

\item \emph{Let }$G=\left(  \left(  T,\lambda\right)  ,A,u,\left(  \Pi
,\pi\right)  ,f\right)  $\emph{ be a nonatomic game with estimation feedback,
}$\delta>0$\emph{, and }$\varepsilon\geq0$\emph{. If }$\lambda$\emph{ is
strongly continuous, }$u=\left(  u_{t}\right)  _{t\in T}$\emph{\ is a family
of functions which is equicontinuous with respect to the second
argument,\footnote{\label{fot:equ}We say that $u=\left(  u_{t}\right)  _{t\in
T}$ is a family of functions which is equicontinuous with respect to the
second argument if and only if\ for each $\varepsilon>0$ there exists
$\delta_{\varepsilon}>0$ such that%
\[
d_{\Delta}\left(  x,y\right)  <\delta_{\varepsilon}\implies\left\vert
u_{t}\left(  a,x\right)  -u_{t}\left(  a,y\right)  \right\vert <\varepsilon
\qquad\forall t\in T,\forall a\in A
\]
}}\ \emph{and each }$f_{t}$\emph{ satisfies condition (\ref{eq:gro})},
\emph{then }$G$\emph{ has a rationalizable\ }$\left(  \delta,\varepsilon
\right)  $\emph{-estimated equilibrium.\footnote{In light of point 1 and,
given the equicontinuity of $u$ (cf. the proof of Corollary \ref{cor:eps-Nas}%
), we can prove that there exists a strategy profile $\sigma\in\Sigma$ such
that
\[
u_{t}\left(  \sigma\left(  t\right)  ,\lambda_{\sigma}^{\pi\left(  t\right)
}\right)  \geq u_{t}\left(  a,\lambda_{\sigma}^{\pi\left(  t\right)  }\right)
-\delta\quad\forall a\in A,\forall t\in T
\]
Set $\tilde{\beta}\in\Delta\left(  S_{0}\right)  ^{T}=\Delta\left(
\Sigma\right)  ^{T}$\ to be such that $\tilde{\beta}\left(  t\right)  $
coincides with the Dirac at $\sigma$ for all $t\in T$.\ It follows that
$\bar{\beta}\left(  t\right)  =\lambda_{\sigma}^{\pi\left(  t\right)  }$ for
all $t\in T$. Given that each $f_{t}$ satisfies (\ref{eq:gro}), we have that
$f_{t}\left(  \sigma\left(  t\right)  ,\bar{\beta}\left(  t\right)
,\lambda_{\sigma}^{\pi\left(  t\right)  }\right)  =0\leq\varepsilon$ for all
$t\in T$. This yields that $\sigma\in S_{1}$. By induction, we can conclude
that $\sigma\in S_{k}$ for all $k\in%
\mathbb{N}
_{0}$, proving that $\sigma$ is a rationalizable\ $\left(  \delta
,\varepsilon\right)  $-estimated equilibrium. The complete\ proof is available
upon request.}}\hfill$\blacktriangle$
\end{enumerate}
\end{remark}

\subsection{Self-confirming and Nash equilibria\label{sec:msg-gam}}

An interesting class of nonatomic games with estimation feedback arises when
the feedback function of player $t$ is generated by a message function
$m_{t}:A\times\Delta\rightarrow M$, where $M$ is a metric space with distance
$d$.\footnote{To simplify notation, we assume that the message space is the
same for all players. This is without loss of generality. We could have
equivalently assumed that each player has his own message space $M_{t}$, and
in the proofs embed this set into a larger common message space $M$. Our
assumptions of equicontinuity on the message functions $m_{t}$ (cf. Corollary
\ref{cor:eps-sce}) would seamlessly pass through the embedding as well.} For
each $t$ in $T$, $m_{t}\left(  a,x\right)  $ represents the \emph{message}
player $t$ receives when he chooses strategy $a$ and the distribution of
players' strategies is $x$. In games with finitely many players, typically the
message function $m_{t}$ depends on the action chosen by the player and the
profile of actions chosen by the opponents. Nevertheless, given our underlying
assumption of anonymity, it seems natural to replace the latter with the
actions' distribution in the population.

With this in mind, the next type of equilibrium models a situation in which
the belief $\beta_{t}$ adopted by each agent $t$\ in equilibrium is
consistent/confirmed\ with/by the message received. More formally, $\beta_{t}$
is such that the expected message $m_{t}\left(  \sigma\left(  t\right)
,\beta_{t}\right)  $ is $\varepsilon$-close to the received message
$m_{t}\left(  \sigma\left(  t\right)  ,\lambda_{\sigma}\right)  $.

We define a \emph{nonatomic game with message feedback} to be a quartet
$G=\left(  \left(  T,\lambda\right)  ,A,u,m\right)  $ where $\left(  \left(
T,\lambda\right)  ,A,u\right)  $ is a nonatomic game and $m=\left(
m_{t}\right)  _{t\in T}$\ is a profile of message functions. Note that a
nonatomic game with \textit{message} feedback can be mapped into a nonatomic
game with \textit{estimation }feedback. In fact, it is enough to consider
$\left(  \Pi,\pi\right)  $ to be trivial, that is $\Pi=\left\{  T\right\}
$,\ and set the profile of feedback functions to be such that:\footnote{In
this case, note that $\pi$ can only take one value.}%
\begin{equation}
f_{t}\left(  a,x,y\right)  =d\left(  m_{t}\left(  a,x\right)  ,m_{t}\left(
a,y\right)  \right)  \quad\forall t\in T,\forall a\in A,\forall x,y\in
\Delta\label{eq:fed-mes}%
\end{equation}
It can be seen immediately that each $f_{t}$ satisfies (\ref{eq:gro}), and
thus (\ref{eq:wea-gro}). We can define our concept\ of self-confirming
$\varepsilon$-equilibrium which we discuss below.

\begin{definition}
Let $\varepsilon\geq0$. \emph{A self-confirming }$\varepsilon$%
\emph{-equilibrium (in pure strategies) }for the nonatomic game with
message\ feedback $G=\left(  \left(  T,\lambda\right)  ,A,u,m\right)  $ is a
strategy profile $\sigma\in\Sigma$ such that there exists a profile of beliefs
$\beta\in\Delta^{T}$ satisfying%
\begin{equation}
\lambda\left(  \left\{  t\in T\ \left\vert \
\begin{array}
[c]{l}%
u_{t}\left(  \sigma\left(  t\right)  ,\beta\left(  t\right)  \right)  \geq
u_{t}\left(  a,\beta\left(  t\right)  \right)  \quad\forall a\in A\\
d\left(  m_{t}\left(  \sigma\left(  t\right)  ,\beta\left(  t\right)  \right)
,m_{t}\left(  \sigma\left(  t\right)  ,\lambda_{\sigma}\right)  \right)
\leq\varepsilon
\end{array}
\right.  \right\}  \right)  =1 \label{eq:e-sce}%
\end{equation}

\end{definition}

In words, a strategy profile $\sigma\in\Sigma$ is a self-confirming
$\varepsilon$-equilibrium ($\varepsilon$-SCE) if and only if

\begin{enumerate}
\item Almost all players best-respond to their beliefs (optimality);

\item Beliefs are not significantly refuted by what they can\ observe
($\varepsilon$-confirmation).
\end{enumerate}

As noted in the Introduction, self-confirming equilibria were introduced for
games with finitely many players by Battigalli \cite{Bat} and Fudenberg and
Levine \cite{FL}, and also $\varepsilon$-confirmation was\ introduced by
Battigalli \cite{Bat}\ and Kalai and Lehrer \cite{KL} and \cite{KL1}. To the
best of our knowledge, the above definition\ of $\varepsilon$-equilibrium
seems to be novel for nonatomic games and also natural (cf. the related
literature section). Furthermore, it encompasses the notions of
self-confirming equilibrium and $\varepsilon$-Nash equilibrium (a fortiori,
Nash equilibrium). To see this, we begin by observing that if $\varepsilon=0$
and $m_{t}:A\times\Delta\rightarrow\Delta$\ is such that%
\begin{equation}
m_{t}\left(  a,x\right)  =x\qquad\forall t\in T,\forall a\in A,\forall
x\in\Delta\label{eq:per-fed}%
\end{equation}
that is, $\left(  M,d\right)  =\left(  \Delta,d_{\Delta}\right)  $ and
feedback is (statistically)\textit{\ perfect}, then (\ref{eq:e-sce}) becomes%
\[
\lambda\left(  \left\{  t\in T\ \left\vert \ u_{t}\left(  \sigma\left(
t\right)  ,\lambda_{\sigma}\right)  \geq u_{t}\left(  a,\lambda_{\sigma
}\right)  \quad\forall a\in A\right.  \right\}  \right)  =1
\]
which means that $\sigma$ is a Nash equilibrium. In this case, beliefs are not
only \textit{perfectly} consistent with observations but also correct.
Maintaining the perfect feedback assumption (\ref{eq:per-fed}), but allowing
for $\varepsilon>0$, (\ref{eq:e-sce}) becomes%
\[
\lambda\left(  \left\{  t\in T\ \left\vert \
\begin{array}
[c]{l}%
u_{t}\left(  \sigma\left(  t\right)  ,\beta\left(  t\right)  \right)  \geq
u_{t}\left(  a,\beta\left(  t\right)  \right)  \quad\forall a\in A\\
d_{\Delta}\left(  \beta\left(  t\right)  ,\lambda_{\sigma}\right)
\leq\varepsilon
\end{array}
\right.  \right\}  \right)  =1
\]
Under a suitable assumption of continuity of $u$ (see Corollary
\ref{cor:eps-Nas} and its proof), we can show that $\sigma$ is an
$\varepsilon$-Nash equilibrium for some suitable $\hat{\varepsilon}>0$, that
is,%
\[
\lambda\left(  \left\{  t\in T\mid u_{t}\left(  \sigma\left(  t\right)
,\lambda_{\sigma}\right)  \geq u_{t}\left(  a,\lambda_{\sigma}\right)
-\hat{\varepsilon}\quad\forall a\in A\right\}  \right)  =1
\]
The intuition is simple: if beliefs are \textquotedblleft
close\textquotedblright\ to the true distribution, players are not far from
objective maximization.

Finally, if we remove perfect feedback\ but maintain\ $\varepsilon=0$,
(\ref{eq:e-sce}) becomes%
\[
\lambda\left(  \left\{  t\in T\ \left\vert \
\begin{array}
[c]{l}%
u_{t}\left(  \sigma\left(  t\right)  ,\beta\left(  t\right)  \right)  \geq
u_{t}\left(  a,\beta\left(  t\right)  \right)  \quad\forall a\in A\\
m_{t}\left(  \sigma\left(  t\right)  ,\beta\left(  t\right)  \right)
=m_{t}\left(  \sigma\left(  t\right)  ,\lambda_{\sigma}\right)
\end{array}
\right.  \right\}  \right)  =1
\]
which is arguably the nonatomic anonymous games counterpart of the definition
of self-confirming equilibrium\ (SCE).

\bigskip

Starting with $\varepsilon$-estimated equilibria, most of our analysis deals
with the case in which $\varepsilon>0$. There are two reasons why we do so.
First, conceptually, $\varepsilon>0$ allows beliefs to be only imperfectly
confirmed, mirroring the fact that players' observations might be noisy and
learning slow. Second, self-confirming equilibria and Nash equilibria might
not exist, as the following examples show. In a nutshell, Example
\ref{exa:KPQS} provides an instance where Nash and SCE equilibria do not
exist, but their $\varepsilon$-versions do. Example \ref{exa:AlN} provides an
instance where $\varepsilon$-uniform Nash equilibria \emph{\`{a} la} Al-Najjar
\cite{AlN}\ do not exist, but standard $\varepsilon$-Nash equilibria do.

\begin{example}
\label{exa:KPQS}\emph{The next example builds on Khan, Qiao, Rath, and Sun
\cite{KQRS}.\footnote{The example of Khan, Qiao, Rath, and Sun \cite{KQRS}
seems to be the first one in the literature to exhibit\ a well-behaved
nonatomic game which does not have any Nash equilibrium, be it pure or
mixed.}\ Consider }$T=%
\mathbb{N}
$\emph{\ and let }$\lambda$\emph{\ be a natural density. Consider two
strategies, that is, }$A=\left\{  1,2\right\}  $\emph{. Assume that for each
}$t\in T$%
\[
u_{t}\left(  a,x\right)  =\left\{
\begin{array}
[c]{cc}%
\frac{1}{t}-x_{1} & a=1\\
x_{1}-\frac{1}{t} & a=2
\end{array}
\right.  \qquad\forall x\in\Delta
\]
\emph{Let }$m_{t}=u_{t}$\emph{\ for all }$t\in T$\emph{. This amounts to the
standard assumption of mere payoff observability.\ Assume that }$\sigma
\in\Sigma$\emph{\ is an\ SCE, that is, there exists }$\beta\in\Delta^{T}%
$\emph{\ such that}%
\[
\lambda\left(  \left\{  t\in T\mid u_{t}\left(  \sigma\left(  t\right)
,\lambda_{\sigma}\right)  =u_{t}\left(  \sigma\left(  t\right)  ,\beta\left(
t\right)  \right)  \geq u_{t}\left(  a,\beta\left(  t\right)  \right)
\quad\forall a\in A\right\}  \right)  =1
\]
\emph{For ease of notation, set }$\lambda_{\sigma}=x$\emph{\ and define the
set of \textquotedblleft optimizing\textquotedblright\ players by}%
\[
O=\left\{  t\in T\mid u_{t}\left(  \sigma\left(  t\right)  ,\lambda_{\sigma
}\right)  =u_{t}\left(  \sigma\left(  t\right)  ,\beta\left(  t\right)
\right)  \geq u_{t}\left(  a,\beta\left(  t\right)  \right)  \quad\forall a\in
A\right\}
\]
\emph{We have two cases:}

\begin{enumerate}
\item $x_{1}>0$\emph{. Since }$\lambda$\emph{\ is a natural density and\ }%
$O$\emph{\ has mass }$1$\emph{, then }$O$\emph{\ is infinite. Thus, there
exists }$\bar{t}\in%
\mathbb{N}
$\emph{\ such that }$\frac{2}{t}-x_{1}<0$\emph{\ for all }$t\in O\cap\left\{
1,...,\bar{t}\right\}  ^{c}$\emph{. Consider }$t\in O\cap\left\{
1,...,\bar{t}\right\}  ^{c}\not =\varnothing$\emph{. By contradiction, assume
that }$\sigma\left(  t\right)  =1$\emph{. The SCE conditions imply that}%
\[
\frac{1}{t}-x_{1}=\frac{1}{t}-\beta\left(  t\right)  _{1}\geq\beta\left(
t\right)  _{1}-\frac{1}{t}%
\]
\emph{yielding that }$0\leq\beta\left(  t\right)  _{1}\leq\frac{2}{t}-x_{1}%
<0$\emph{, a contradiction. Since }$t$\emph{\ was arbitrarily chosen in
}$O\cap\left\{  1,...,\bar{t}\right\}  ^{c}$\emph{, it follows that }%
$\sigma\left(  t\right)  =2$\emph{\ for all }$t\in O\cap\left\{  1,...,\bar
{t}\right\}  ^{c}$\emph{. Since }$\lambda$\emph{\ is a natural density and
}$O$\emph{\ and }$O\cap\left\{  1,...,\bar{t}\right\}  ^{c}$\emph{\ differ by
a finite set }$\lambda\left(  O\cap\left\{  1,...,\bar{t}\right\}
^{c}\right)  =1$\emph{, we have that }$\lambda_{\sigma}=x$\emph{\ is such that
}$x_{2}=1$\emph{, a contradiction with }$0=1-x_{2}=x_{1}>0$\emph{.}

\item $x_{1}=0$\emph{. Consider }$t\in O$\emph{. By contradiction, assume that
}$\sigma\left(  t\right)  =2$\emph{. The SCE conditions imply that}%
\[
x_{1}-\frac{1}{t}=\beta\left(  t\right)  _{1}-\frac{1}{t}\geq\frac{1}{t}%
-\beta\left(  t\right)  _{1}%
\]
\emph{yielding that }$0=x_{1}=\beta\left(  t\right)  _{1}$\emph{\ and }%
$0\geq\frac{2}{t}>0$\emph{, a contradiction. Since }$t$\emph{\ was arbitrarily
chosen in }$O$, $\sigma\left(  t\right)  =1$\emph{\ for all }$t\in O$\emph{,
yielding that }$\lambda_{\sigma}=x$\emph{\ is such that }$x_{1}=1$\emph{, a
contradiction with }$x_{1}=0$\emph{.}
\end{enumerate}

\noindent\emph{To sum up, we have just shown that the nonatomic game with
message feedback above does not have any self-confirming equilibrium and, in
particular, any Nash equilibrium.}\footnote{Two extra observations are in
order:
\par
a. In the nonatomic game above, SCE equilibria and Nash equilibria coincide.
This is by chance, as the\ next point shows.
\par
b. Khan, Qiao, Rath, and Sun \cite{KQRS} consider $T=%
\mathbb{N}
$ and let $\lambda$\ be a natural density. They assume\ $A=\left\{
1,2\right\}  $ and $\hat{u}$ to be such that for each $t\in T$%
\[
\hat{u}_{t}\left(  a,x\right)  =\left\{
\begin{array}
[c]{cc}%
\frac{1}{t}-x_{1} & a=1\\
0 & a=2
\end{array}
\right.  \quad\forall x\in\Delta
\]
With similar arguments, they prove that the nonatomic game $\left(  \left(
T,\lambda\right)  ,A,\hat{u}\right)  $ does not have any Nash equilibrium.\ At
the same time, if we consider the augmented nonatomic game with message
feedback $\left(  \left(  T,\lambda\right)  ,A,\hat{u},m\right)  $\ where
$m_{t}=\hat{u}_{t}$\ for all $t\in T$, then we can show that there exists an
SCE equilibrium. In fact, if $\sigma\in\Sigma$ is such that $\sigma\left(
t\right)  =2$ for all $t\in T$, by setting $\beta\in\Delta^{T}$ such that
$\beta\left(  t\right)  _{1}=1$ for all $t\in T$, we obtain the result.}%
\emph{\ This happens despite the fact that the profile of message functions is
extremely well-behaved being }$m=\left(  m_{t}\right)  _{t\in T}%
$\emph{\ equicontinuous with respect to the second argument (cf. Corollary
\ref{cor:eps-sce}).}\footnote{Indeed, note that for each $\varepsilon>0$ we
can set $\delta_{\varepsilon}=\varepsilon$ and get%
\[
d_{\Delta}\left(  x,y\right)  <\varepsilon\implies\left\vert m_{t}\left(
a,x\right)  -m_{t}\left(  a,y\right)  \right\vert =\left\vert x_{1}%
-y_{1}\right\vert \leq d_{\Delta}\left(  x,y\right)  <\varepsilon\quad\forall
t\in T,\forall a\in A
\]
}\emph{\ At the same time, consider }$\varepsilon>0$\emph{.\ Let }$\bar{t}\in%
\mathbb{N}
$\emph{\ be such that }$\min\left\{  1,\varepsilon\right\}  >\frac{1}{t}$
\emph{for all }$t\in%
\mathbb{N}
$\emph{\ such that\ }$t>\bar{t}$\emph{. Set }$\bar{\varepsilon}=\min\left\{
1,\varepsilon\right\}  $\emph{. Consider a strategy profile }$\sigma\in\Sigma
$\emph{\ and a belief profile }$\beta\in\Delta^{T}\ $\emph{such that }%
$\sigma\left(  t\right)  =2$\emph{\ and }$\beta\left(  t\right)  _{1}%
=\frac{\bar{\varepsilon}+\frac{1}{t}}{2}\in\left(  \frac{1}{t},\bar
{\varepsilon}\right)  \subseteq\left(  0,1\right)  $\emph{\ for all }$t\in%
\mathbb{N}
$\emph{\ such that\ }$t>\bar{t}$\emph{. Since }$\left\{  1,...,\bar
{t}\right\}  $\emph{\ is finite and }$\lambda$\emph{\ is a natural density,}
\emph{we have that }$\lambda_{\sigma}=x$\emph{\ is such that }$x_{2}=1$\emph{,
that is, }$x_{1}=0$\emph{. It follows that for each }$t\in\left\{
1,...,\bar{t}\right\}  ^{c}$%
\[
\left\vert m_{t}\left(  \sigma\left(  t\right)  ,\beta\left(  t\right)
\right)  -m_{t}\left(  \sigma\left(  t\right)  ,\lambda_{\sigma}\right)
\right\vert =\left\vert \frac{\bar{\varepsilon}+\frac{1}{t}}{2}-\frac{1}%
{t}-x_{1}+\frac{1}{t}\right\vert =\frac{\bar{\varepsilon}+\frac{1}{t}}{2}%
<\bar{\varepsilon}\leq\varepsilon
\]
\emph{and}%
\[
u_{t}\left(  \sigma\left(  t\right)  ,\beta\left(  t\right)  \right)
=\beta\left(  t\right)  _{1}-\frac{1}{t}=\frac{\bar{\varepsilon}-\frac{1}{t}%
}{2}>0>\frac{1}{t}-\beta\left(  t\right)  _{1}=u_{t}\left(  1,\beta\left(
t\right)  \right)
\]
\emph{Since }$\left\{  1,...,\bar{t}\right\}  ^{c}$\emph{\ has mass }%
$1$\emph{,\ we can conclude that }$\sigma\in\Sigma$\emph{\ is an }%
$\varepsilon$\emph{-SCE.}\hfill$\blacktriangle$
\end{example}

\begin{example}
\label{exa:AlN}\emph{Al-Najjar \cite{AlN}\ (cf. the Introduction) also deals
with the lack of countable additivity and studies the following equilibrium: a
strategy }$\sigma\in\Sigma$\emph{ is an Al-Najjar equilibrium (in pure
strategies) if and only if for each }$\varepsilon>0$%
\begin{equation}
\lambda\left(  \left\{  t\in T\mid u_{t}\left(  \sigma\left(  t\right)
,\lambda_{\sigma}\right)  \geq u_{t}\left(  a,\lambda_{\sigma}\right)
-\varepsilon\quad\forall a\in A\right\}  \right)  >1-\varepsilon
\label{eq:exa-equ}%
\end{equation}
\emph{We next show that also these equilibria might fail to exist. In what
follows, it will often be useful to set}%
\[
O_{\varepsilon}=\left\{  t\in T\mid u_{t}\left(  \sigma\left(  t\right)
,\lambda_{\sigma}\right)  \geq u_{t}\left(  a,\lambda_{\sigma}\right)
-\varepsilon\quad\forall a\in A\right\}
\]
\emph{Two observations are in order. First, compared to the }$\varepsilon
$\emph{-Nash equilibria we study (Corollary \ref{cor:eps-Nas}), the key
difference is that, in our case, }$\sigma$\emph{ might depend on the given
}$\varepsilon$\emph{, while in Al-Najjar's case, }$\sigma$\emph{ must work
with any }$\varepsilon$\emph{. In particular, one can easily show
that:\footnote{It is easy to see that if $0<\varepsilon<\varepsilon^{\prime}$,
then $O_{\varepsilon}\subseteq O_{\varepsilon^{\prime}}$, thus%
\[
\lambda\left(  O_{\varepsilon^{\prime}}\right)  \geq\lambda\left(
O_{\varepsilon}\right)  >1-\varepsilon\quad\forall\varepsilon^{\prime
}>0,\forall\varepsilon\in\left(  0,\varepsilon^{\prime}\right)
\]
yielding that $\lambda\left(  O_{\varepsilon^{\prime}}\right)  =1$ for all
$\varepsilon^{\prime}>0$.} }$\sigma\in\Sigma$\emph{ is\ an Al-Najjar
equilibrium if and only if for each }$\varepsilon>0$%
\[
\lambda\left(  \left\{  t\in T\mid u_{t}\left(  \sigma\left(  t\right)
,\lambda_{\sigma}\right)  \geq u_{t}\left(  a,\lambda_{\sigma}\right)
-\varepsilon\quad\forall a\in A\right\}  \right)  =1
\]
\emph{Second, by taking the intersection of the sets }$O_{1/n}$\emph{, this
allows us to conclude easily that an Al-Najjar equilibrium is a Nash
equilibrium, provided\ }$\lambda$\emph{ is countably additive. We consider the
nonatomic game }$\left(  \left(  T,\lambda\right)  ,A,\tilde{u}\right)
$\emph{ where }$\left(  T,\lambda\right)  \ $\emph{and }$A$\emph{\ are as in
Example \ref{exa:KPQS}\ and for each }$t\in T$%
\[
\tilde{u}_{t}\left(  a,x\right)  =\left\{
\begin{array}
[c]{cc}%
\frac{1}{t}-x_{1} & a=1\text{\emph{ and }}x_{1}>0\\
x_{1}-\frac{1}{t} & a=2\text{\emph{ and }}x_{1}>0\\
1 & a=1\text{\emph{ and }}x_{1}=0\\
\frac{1}{t} & a=2\text{\emph{ and }}x_{1}=0
\end{array}
\right.  \qquad\forall x\in\Delta
\]
\emph{Assume that }$\sigma\in\Sigma$\emph{\ satisfies (\ref{eq:exa-equ}). For
ease of notation, set }$\lambda_{\sigma}=x$\emph{. As before, we have two
cases:}

\begin{enumerate}
\item $x_{1}>0$\emph{. Fix }$\varepsilon>0$\emph{. Since }$\lambda$\emph{\ is
a natural density, the set }$O_{\varepsilon}$\emph{\ has mass }$1$\emph{, and
}$\lambda\left(  \left\{  t\in T\mid\sigma\left(  t\right)  =1\right\}
\right)  >0$\emph{, we have that\ }$O_{\varepsilon}\cap\left\{  t\in
T\mid\sigma\left(  t\right)  =1\right\}  $\emph{\ is infinite. Since
}$\varepsilon$\emph{ was arbitrarily chosen, this implies that we can
construct a strictly increasing\ sequence }$\left\{  t_{k}\right\}  _{k\in%
\mathbb{N}
}\subseteq%
\mathbb{N}
$\emph{\ such that }$t_{k}\in O_{1/k}\cap\left\{  t\in T\mid\sigma\left(
t\right)  =1\right\}  $\emph{ for all }$k\in%
\mathbb{N}
$\emph{. Since }$t_{k}\in O_{1/k}$\emph{, }$\sigma\left(  t_{k}\right)
=1$\emph{, and }$x_{1}>0$\emph{, we have that for each }$k\in%
\mathbb{N}
$%
\[
\frac{1}{t_{k}}-x_{1}\geq x_{1}-\frac{1}{t_{k}}-\frac{1}{k}\implies0<x_{1}%
\leq\frac{1}{t_{k}}+\frac{1}{2k}%
\]
\emph{By passing to the limit, we obtain that }$0<x_{1}\leq0$\emph{, a
contradiction.}

\item $x_{1}=0$\emph{. Fix }$\varepsilon>0$\emph{. Since }$\lambda$\emph{\ is
a natural density, the set }$O_{\varepsilon}$\emph{\ has mass }$1$\emph{, and
}$\lambda\left(  \left\{  t\in T\mid\sigma\left(  t\right)  =2\right\}
\right)  >0$\emph{, we have that\ }$O_{\varepsilon}\cap\left\{  t\in
T\mid\sigma\left(  t\right)  =2\right\}  $\emph{\ is infinite. Since
}$\varepsilon$\emph{ was arbitrarily chosen, this implies that we can
construct a strictly increasing\ sequence }$\left\{  t_{k}\right\}  _{k\in%
\mathbb{N}
}\subseteq%
\mathbb{N}
$\emph{\ such that }$t_{k}\in O_{1/k}\cap\left\{  t\in T\mid\sigma\left(
t\right)  =2\right\}  $\emph{ for all }$k\in%
\mathbb{N}
$\emph{. Since }$t_{k}\in O_{1/k}$\emph{, }$\sigma\left(  t_{k}\right)
=2$\emph{, and }$x_{1}=0$\emph{, we have that for each }$k\in%
\mathbb{N}
$%
\[
\frac{1}{t_{k}}\geq1-\frac{1}{k}%
\]
\emph{By passing to the limit, we obtain that }$0\geq1$\emph{, a
contradiction.}
\end{enumerate}

\noindent\emph{To sum up, we have just shown that the nonatomic game }$\left(
\left(  T,\lambda\right)  ,A,\tilde{u}\right)  $\emph{ does not have any
equilibrium as defined in (\ref{eq:exa-equ}). At the same time, it is not hard
to see that this game admits an }$\varepsilon$\emph{-Nash equilibrium for
every }$\varepsilon>0$\emph{. One way to observe this is to consider the
augmented game }$\left(  \left(  T,\lambda\right)  ,A,\tilde{u},m\right)  $
\emph{in which each player has perfect statistical feedback: that is}%
\[
m_{t}\left(  a,x\right)  =x\qquad\forall t\in T,\forall a\in A,\forall
x\in\Delta
\]
\emph{Since\ }$m=\left(  m_{t}\right)  _{t\in T}$\emph{\ is equicontinuous
with respect to the second argument (cf. Corollary \ref{cor:eps-sce}), we have
that for each }$\varepsilon>0$\emph{\ there exists an }$\varepsilon
$\emph{-SCE. Given }$\varepsilon\in\left(  0,1\right)  $\emph{, it can
immediately be proved that a strategy profile }$\sigma$\ \emph{is an
}$\varepsilon$\emph{-SCE if and only if }$\lambda\left(  \left\{  t\in
T\mid\sigma\left(  t\right)  =1\right\}  \right)  \in\left[  0,\varepsilon
/\sqrt{2}\right]  $\emph{. Given our choice of }$m$\emph{, following the
intuition that \textquotedblleft if beliefs are close\ to the true
distribution, players are not far from objective
maximization\textquotedblright, we can prove that, given }$\varepsilon
\in\left(  0,1\right)  $\emph{,\ if }$\sigma$\ \emph{is an }$\frac
{\varepsilon}{2\sqrt{2}}$\emph{-SCE and }$\lambda\left(  \left\{  t\in
T\mid\sigma\left(  t\right)  =1\right\}  \right)  >0$\emph{, then }$\sigma
$\emph{ is an }$\varepsilon$\emph{-Nash equilibrium. In other words, }$\left(
\left(  T,\lambda\right)  ,A,\tilde{u}\right)  $\emph{ does not have any
equilibrium as defined in (\ref{eq:exa-equ}), but for each }$\varepsilon
>0$\emph{ it has an }$\varepsilon$\emph{-Nash equilibrium.\footnote{Since
$\lambda$ is strongly continuous, note that, given $\varepsilon\in\left(
0,1\right)  $, we can always find $\sigma\in\Sigma$ such that $\lambda\left(
\left\{  t\in T\mid\sigma\left(  t\right)  =1\right\}  \right)  \in\left(
0,\varepsilon/4\right]  $. In other words, in light of the above
characterization, we can always find a strategy profile $\sigma$ which is an
$\frac{\varepsilon}{2\sqrt{2}}$-SCE such that $\lambda\left(  \left\{  t\in
T\mid\sigma\left(  t\right)  =1\right\}  \right)  >0$.}}\hfill$\blacktriangle$
\end{example}

We are ready to state the\ main results of\ this section.

\begin{corollary}
\label{cor:eps-sce}Let $G=\left(  \left(  T,\lambda\right)  ,A,u,m\right)  $
be a nonatomic game with message feedback and $\varepsilon>0$.\ If $\lambda$
is strongly continuous\ and $m=\left(  m_{t}\right)  _{t\in T}$\ is a family
of functions which is equicontinuous with respect to the second
argument,\footnote{We say that $m=\left(  m_{t}\right)  _{t\in T}$ is a family
of functions which is equicontinuous with respect to the second argument if
and only if\ for each $\varepsilon>0$ there exists $\delta_{\varepsilon}>0$
such that%
\[
d_{\Delta}\left(  x,y\right)  <\delta_{\varepsilon}\implies d\left(
m_{t}\left(  a,x\right)  ,m_{t}\left(  a,y\right)  \right)  <\varepsilon
\qquad\forall t\in T,\forall a\in A
\]
} then $G$ has an $\varepsilon$-SCE.
\end{corollary}

In particular, under the assumption of payoff observability, that is,
$m_{t}\left(  a,x\right)  =u_{t}\left(  a,x\right)  $ for all $t\in T,a\in
A,x\in\Delta$,\ Corollary \ref{cor:eps-sce}\ yields\ that if $\lambda$ is
strongly continuous and $u=\left(  u_{t}\right)  _{t\in T}$\ is a family of
functions which is equicontinuous with respect to the second argument, then
there exists an $\varepsilon$-SCE strategy profile $\sigma\in\Sigma$ such that%
\[
\lambda\left(  \left\{  t\in T\ \left\vert \
\begin{array}
[c]{l}%
u_{t}\left(  \sigma\left(  t\right)  ,\beta\left(  t\right)  \right)  \geq
u_{t}\left(  a,\beta\left(  t\right)  \right)  \quad\forall a\in A\\
\left\vert u_{t}\left(  \sigma\left(  t\right)  ,\beta\left(  t\right)
\right)  -u_{t}\left(  \sigma\left(  t\right)  ,\lambda_{\sigma}\right)
\right\vert \leq\varepsilon
\end{array}
\right.  \right\}  \right)  =1
\]
where $\beta\in\Delta^{T}$. In this case, the objective observed payoff
substantially matches the expected one.\ Building on Corollary
\ref{cor:eps-sce} and following a similar intuition,\ we also obtain the
existence of $\varepsilon$-Nash equilibria.

\begin{corollary}
\label{cor:eps-Nas}Let $G=\left(  \left(  T,\lambda\right)  ,A,u\right)  $ be
a nonatomic game and $\varepsilon>0$.\ If $\lambda$ is strongly
continuous\ and $u=\left(  u_{t}\right)  _{t\in T}$\ is a family of functions
which is equicontinuous with respect to the second argument,\footnote{See
Footnote \ref{fot:equ}.}\ then $G$ has an $\varepsilon$-Nash equilibrium, that
is, there exists a strategy profile $\sigma\in\Sigma$ such that%
\[
\lambda\left(  \left\{  t\in T\ \left\vert \ u_{t}\left(  \sigma\left(
t\right)  ,\lambda_{\sigma}\right)  \geq u_{t}\left(  a,\lambda_{\sigma
}\right)  -\varepsilon\quad\forall a\in A\right.  \right\}  \right)  =1
\]

\end{corollary}

At this point, the reader might wonder how restrictive are our assumptions
of\ equicontinuity.\footnote{See also the discussion following Corollary
\ref{cor:eps-pce}.} At first sight, it might appear that the degree of
similarity among players, imposed by a measurable structure as in the original
framework of Schmeidler, is here replaced by equicontinuity. The following
example should clarify that this is far from being the case.

\begin{example}
\label{exa:equ-con-u}\emph{Assume that players have expected-utility like
preferences, namely, for each }$t\in T$%
\[
u_{t}\left(  a,x\right)  =\sum_{b\in A}v_{t}\left(  a,b\right)  x_{b}%
\qquad\forall a\in A,\forall x\in\Delta
\]
\emph{where }$v_{t}:A\times A\rightarrow%
\mathbb{R}
$\emph{. As is well known, each\ }$v_{t}$\emph{\ can be normalized to be
taking values in }$\left[  0,1\right]  $\emph{, without altering the player's
preferences. In light of this, an immediate application of the Cauchy-Schwarz
inequality yields that}%
\[
\left\vert u_{t}\left(  a,x\right)  -u_{t}\left(  a,y\right)  \right\vert
\leq\sqrt{n}d_{\Delta}\left(  x,y\right)  \qquad\forall t\in T,\forall a\in A
\]
\emph{proving that }$u=\left(  u_{t}\right)  _{t\in T}$\ \emph{is a family of
functions which is equicontinuous with respect to the second argument. Thus,
preferences can be extremely different within the above class and yet satisfy
equicontinuity.}\hfill$\blacktriangle$
\end{example}

As mentioned in the Introduction, Khan, Qiao, Rath, and Sun \cite{KQRS} showed
that in the absence of countable additivity the existence of Nash equilibria
is not guaranteed. They also reported\ an independent result of existence of
an\ $\varepsilon$-Nash equilibrium. Their definition is weaker than ours. In
their case, a strategy profile $\sigma\in\Sigma$ is an $\varepsilon
$-equilibrium if and only if%
\[
\lambda\left(  \left\{  t\in T\ \left\vert \ u_{t}\left(  \sigma\left(
t\right)  ,\lambda_{\sigma}\right)  \geq u_{t}\left(  a,\lambda_{\sigma
}\right)  -\varepsilon\quad\forall a\in A\right.  \right\}  \right)
\geq1-\varepsilon
\]

\subsection{Peer-confirming equilibria\label{sec:nei-gam}}

Lipnowski\ and Sadler \cite{LS}\ propose a\ notion of equilibrium in which
players\ best-respond to beliefs which are required to be correct only when it
comes to the behavior of opponents who belong to the same neighborhood.
Moreover, they further require that this is correctly and commonly believed by
players. Formally, the collection\ of neighborhoods is a partition of the
players in terms of the connected components of an underlying undirected
network. They study games with finitely many players. For simultaneous-move
games, peer-confirming equilibrium is an example of rationalizable
self-confirming equilibrium (see also Rubinstein and Wolinsky \cite{RW}\ as
well as Fudenberg and Kamada \cite{FK}). In what follows, we dispense with the
assumption of correct and common belief. This seems reasonable since nonatomic
anonymous games model exactly situations where individuals are negligible and
are not fully aware of the strategic environment surrounding them,
rendering\ sophisticated strategic reasoning less\ plausible. At the same
time, our notion of rationalizable $\left(  \delta,\varepsilon\right)
$-estimated equilibrium allows us to offer a more faithful translation to our
setting of peer-confirming equilibrium (see Remark \ref{rmk:eps-PCE}%
).\ Moreover, given anonymity we require that players'\ observations are only
about the actions' distributions in the subpopulation they face.

We define a \emph{nonatomic game with a neighborhood structure }to be a
quartet $G=(\left(  T,\lambda\right)  ,A,u,\left(  \Pi,\pi\right)  )$ where
$\left(  \left(  T,\lambda\right)  ,A,u\right)  $ is a nonatomic game and
$\left(  \Pi,\pi\right)  $\ is a neighborhood structure. Note that a nonatomic
game with \textit{a neighborhood structure} can be mapped into a nonatomic
game with \textit{estimation feedback}. In fact, it is enough to set the
profile of feedback functions to be such that%
\begin{equation}
f_{t}\left(  a,x,y\right)  =d_{\Delta}\left(  x,y\right)  \quad\forall t\in
T,\forall a\in A,\forall x,y\in\Delta\label{eq:fed-nei-str}%
\end{equation}
It can be seen immediately that each $f_{t}$ satisfies (\ref{eq:gro}), and
thus (\ref{eq:wea-gro}). We can define the\ version of peer-confirming
$\varepsilon$-equilibrium that\ we analyze below.

\begin{definition}
Let $\varepsilon\geq0$. \emph{A peer-confirming }$\varepsilon$%
\emph{-equilibrium (in pure strategies) }for the nonatomic game with a
neighborhood structure $G=\left(  \left(  T,\lambda\right)  ,A,u,\left(
\Pi,\pi\right)  \right)  $ is a strategy profile $\sigma\in\Sigma$ such that
there exists a profile of beliefs $\beta\in\Delta^{T}$ satisfying%
\begin{equation}
\lambda\left(  \left\{  t\in T\ \left\vert \
\begin{array}
[c]{l}%
u_{t}\left(  \sigma\left(  t\right)  ,\beta\left(  t\right)  \right)  \geq
u_{t}\left(  a,\beta\left(  t\right)  \right)  \quad\forall a\in A\\
d_{\Delta}\left(  \beta\left(  t\right)  ,\lambda_{\sigma}^{\pi\left(
t\right)  }\right)  \leq\varepsilon
\end{array}
\right.  \right\}  \right)  =1 \label{eq:e-pce}%
\end{equation}

\end{definition}

In words, a strategy profile $\sigma\in\Sigma$ is a peer-confirming
$\varepsilon$-equilibrium ($\varepsilon$-PCE) if and only if

\begin{enumerate}
\item Almost all players best-respond\ to their beliefs (optimality);

\item Beliefs are almost correct in terms of the subpopulation observed
($\varepsilon$-neighborhood confirmation).
\end{enumerate}

\begin{corollary}
\label{cor:eps-pce}Let $G=\left(  \left(  T,\lambda\right)  ,A,u,\left(
\Pi,\pi\right)  \right)  $ be a nonatomic game with a neighborhood structure
and $\varepsilon>0$.\ If $\lambda$ is strongly continuous, then $G$ has\ an
$\varepsilon$-PCE.
\end{corollary}

It is important to note how the corollary above does not require any extra
property of continuity. For, in such a case feedback is perfect,
when\ restricted to each subpopulation, and action independent, automatically
satisfying\ the requirement of equicontinuity in Theorem \ref{thm:mai}.
Conceptually, this confirms that, in contrast with measurability assumptions,
our properties of equicontinuity do not impose automatically\ that
\textquotedblleft close players\textquotedblright\ have similar
preferences/behavior\ (cf. Example \ref{exa:equ-con-u}).

\begin{remark}
\label{rmk:eps-PCE}\emph{Two observations are in order:}

\begin{enumerate}
\item \emph{In the definition of }$\varepsilon$\emph{-PCE, we could allow for
the possibility that each player }$t$\emph{\ has a belief }$\tilde{\beta}_{t}%
$\ \emph{over the entire space of players' strategy profiles }$\Sigma$
\emph{(cf. point 2 of Remark \ref{rmk:RSCE})}\ \emph{and require that only the
restriction to the subpopulation observed, in terms of actions' distribution,
that is }$\bar{\beta}_{t}$\emph{,\ is }$\varepsilon$\emph{-confirmed. This
would allow for modelling explicitly the possibility that players, in
equilibrium,\ possibly entertain wrong beliefs about players not in their
neighborhood. Given our nonatomic structure and Corollary \ref{cor:eps-pce},
we could obtain an existence result also for this more general notion.}

\item \emph{Consider a rationalizable }$\left(  \delta,\varepsilon\right)
$\emph{-estimated equilibrium as defined in point 2 of Remark \ref{rmk:RSCE}
where the profile of feedback functions is set to be as in
(\ref{eq:fed-nei-str}). Given this specification, in a rationalizable
}$\left(  \delta,\varepsilon\right)  $\emph{-estimated equilibrium, all
players }$\delta$\emph{-best-respond\ to their beliefs which are almost
correct in terms of the subpopulation observed. Moreover, this is correctly
and commonly believed by all players. By setting }$\varepsilon=\delta
=0$\emph{, our definition provides a more faithful formal translation to
nonatomic anonymous games of the equilibrium notion studied by Lipnowski\ and
Sadler \cite{LS}. By point 3 of Remark \ref{rmk:RSCE}, given a nonatomic game
with a neighborhood structure} $G=\left(  \left(  T,\lambda\right)
,A,u,\left(  \Pi,\pi\right)  \right)  $\emph{ as well as\ }$\delta>0$\emph{
and }$\varepsilon\geq0$\emph{, if }$\lambda$\emph{ is strongly continuous\ and
}$u=\left(  u_{t}\right)  _{t\in T}$\emph{\ is a family of functions which is
equicontinuous with respect to the second argument}, \emph{then }$G$\emph{ has
a rationalizable\ }$\left(  \delta,\varepsilon\right)  $\emph{-estimated
equilibrium.}\hfill$\blacktriangle$
\end{enumerate}
\end{remark}

\subsection{Berk--Nash equilibria\label{sec:mis-gam}}

Esponda and Pouzo \cite{EP}\ propose a\ notion of equilibrium that allows for
players' beliefs to be possibly misspecified (see also Remark
\ref{rmk:eps-BNE} below). It is a different way, compared to self-confirming
equilibria, to allow for potentially incorrect beliefs in equilibrium. They
term their notion of equilibrium Berk--Nash. Berk--Nash equilibria are based
on the assumption that each player has a set of probabilistic models over the
payoff-relevant features, in our case $\left\{  Q_{t}\right\}  _{t\in
T}\subseteq\Delta^{o}$,\footnote{As usual, $\Delta^{o}$ denotes the set%
\[
\left\{  x\in\Delta\mid x_{i}>0\quad\forall i\in\left\{  1,...,n\right\}
\right\}
\]
In other words, $\Delta^{o}$ is the relative interior of $\Delta$.}\ and:

\begin{enumerate}
\item All players best-respond to their beliefs (optimality);

\item Each player's belief is restricted to be the best fit (in terms
of\ Kullback--Leibler distance) among the set of beliefs he considers possible.
\end{enumerate}

In our setup, this would mean that each player $t$ has a (possibly
misspecified) set of models $Q_{t}\subseteq\Delta^{o}$. A\ strategy profile
$\sigma\in\Sigma$ is a Berk--Nash equilibrium\ if and only if there exists a
profile of beliefs $\beta\in\Delta^{T}$ such that the set of all players that
satisfy the following two conditions has full measure:\footnote{Compared to
Esponda and Pouzo \cite{EP}, we do not assume that players' are expected
utility and have a prior $\mu$ over $\operatorname{argmin}_{z\in Q_{t}%
}K\left(  \lambda_{\sigma}||z\right)  $. In other words,\ players are only
allowed to consider degenerate priors. A priori, this makes it more difficult
to obtain an existence result. Moreover, we are also not considering any extra
form of feedback (see point 1 of Remark \ref{rmk:eps-BNE} below).}

\begin{enumerate}
\item $u_{t}\left(  \sigma\left(  t\right)  ,\beta\left(  t\right)  \right)
\geq u_{t}\left(  a,\beta\left(  t\right)  \right)  $ for all$\ a\in A$;

\item $\beta\left(  t\right)  \in\operatorname{argmin}_{z\in Q_{t}}K\left(
\lambda_{\sigma}||z\right)  $ (where $K$ is the Kullback--Leibler distance).
\end{enumerate}

In what follows, we offer a more general version for nonatomic games of the
above equilibrium. In order to do so, we define a \emph{nonatomic game with
model misspecification} to be a quintet\ $G=\left(  \left(  T,\lambda\right)
,A,u,\mathcal{Q},D\right)  $ where

\begin{enumerate}
\item[a.] $\left(  \left(  T,\lambda\right)  ,A,u\right)  $ is a nonatomic game;

\item[b.] $\mathcal{Q}=\left(  Q_{t}\right)  _{t\in T}$\ is a profile of sets
of actions' distributions, that is, $Q_{t}$ is a nonempty, compact, and convex
subset of $\Delta^{o}$ for all $t\in T$;

\item[c.] $D:\Delta\times\Delta^{o}\rightarrow\left[  0,\infty\right)  $ is a
statistical divergence, that is, a jointly convex and continuous function such
that for each\ $x,y\in\Delta^{o}$%
\begin{equation}
x=y\iff D\left(  x||y\right)  =0. \label{eq:dis-zer}%
\end{equation}
The next example describes a class of widely used statistical divergences.
\end{enumerate}

\begin{example}
\label{exa:div}\emph{The most classic statistical divergences\ are }$\phi
$\emph{-divergences which have the form}%
\[
D_{\phi}\left(  x||z\right)  =\sum_{i=1}^{n}z_{i}\phi\left(  \frac{x_{i}%
}{z_{i}}\right)
\]
\emph{where }$\phi$\emph{\ is a positive, continuous, strictly convex function
on }$\mathbb{R}_{+}$\emph{ such that }$\phi\left(  1\right)  =0$\emph{.\ For
example, for }$\phi\left(  s\right)  =s\log s-s+1$\emph{,\footnote{Here, it is
assumed implicitly that $\phi\left(  0\right)  =1$ which is obtained by taking
the limit for $s\rightarrow0$.} }$D_{\phi}$\emph{\ is the Kullback--Leibler
distance, for }$\phi\left(  s\right)  =\left(  s-1\right)  ^{2}/2$\emph{,
}$D_{\phi}$\emph{\ is the }$\chi^{2}$\emph{-distance, and for }$\phi\left(
s\right)  =\left(  \sqrt{s}-1\right)  ^{2}$\emph{, }$D_{\phi}$\emph{\ is the
Hellinger distance. In all these specifications, }$D_{\phi}$\emph{\ satisfies
(\ref{eq:dis-zer}) and it is jointly convex and\ continuous.}\hfill
$\blacktriangle$
\end{example}

Note that a nonatomic game with \textit{model misspecification} can be mapped
into a nonatomic game with \textit{estimation feedback}. In fact, it is enough
to consider $\left(  \Pi,\pi\right)  $ to be trivial, that is $\Pi=\left\{
T\right\}  $, and set the profile of feedback functions to be such
that:\footnote{In this case, note that $\pi$ can only take one value.
Moreover, when $x\in\Delta$ and $Y$ is a nonempty subset of $\Delta$,
$d_{\Delta}\left(  x,Y\right)  $ denotes the distance of $x$ from the set $Y$,
that is,%
\[
d_{\Delta}\left(  x,Y\right)  =\inf_{y\in Y}d_{\Delta}\left(  x,y\right)
\]
In our case, $Y=\operatorname{argmin}_{z\in Q_{t}}D\left(  y||z\right)  $.}%
\begin{equation}
f_{t}\left(  a,x,y\right)  =d_{\Delta}\left(  x,\operatorname{argmin}_{z\in
Q_{t}}D\left(  y||z\right)  \right)  \quad\forall t\in T,\forall a\in
A,\forall x,y\in\Delta\label{eq:pro-div}%
\end{equation}
It is not hard to show that each $f_{t}$ satisfies (\ref{eq:wea-gro}), but
might fail to satisfy (\ref{eq:gro}). We can define our version of Berk--Nash
$\varepsilon$-equilibrium which we discuss below.

\begin{definition}
\label{def:eps-BNE}Let $\varepsilon\geq0$. \emph{A Berk--Nash }$\varepsilon
$\emph{-equilibrium (in pure strategies) }for the nonatomic game with model
misspecification\ $G=\left(  \left(  T,\lambda\right)  ,A,u,\mathcal{Q}%
,D\right)  $ is a strategy profile $\sigma\in\Sigma$ such that there exists a
profile of beliefs $\beta\in\Delta^{T}$ satisfying%
\begin{equation}
\lambda\left(  \left\{  t\in T\ \left\vert \
\begin{array}
[c]{l}%
u_{t}\left(  \sigma\left(  t\right)  ,\beta\left(  t\right)  \right)  \geq
u_{t}\left(  a,\beta\left(  t\right)  \right)  \quad\forall a\in A\\
d_{\Delta}\left(  \beta\left(  t\right)  ,\operatorname{argmin}_{z\in Q_{t}%
}D\left(  \lambda_{\sigma}||z\right)  \right)  \leq\varepsilon
\end{array}
\right.  \right\}  \right)  =1 \label{eq:e-BNE}%
\end{equation}

\end{definition}

Note that a strategy profile $\sigma\in\Sigma$ is a Berk--Nash $\varepsilon
$-equilibrium ($\varepsilon$-BNE) if and only if

\begin{enumerate}
\item Almost all players best-respond to their beliefs (optimality);

\item Beliefs are $\varepsilon$-close to the set of probabilistic models which
are the best fit in the primitive set $Q_{t}$ of the realized distribution
($\varepsilon$-fit).
\end{enumerate}

Although prima facie they might appear similar, the notion of $\varepsilon
$-BNE is conceptually and formally very different from that of $\varepsilon
$-SCE.\footnote{The two equilibrium notions are distinct, but share some
overlap (see Esponda and Pouzo \cite{EP}).} The next result proves that, under
suitable conditions, the former type of equilibria always exists. To do so, we
need a last piece of notation. Given $\delta>0$, we denote%
\[
\Delta_{\delta}=\left\{  x\in\Delta\mid x_{i}\geq\delta\quad\forall
i\in\left\{  1,..,n\right\}  \right\}
\]
In words, $\Delta_{\delta}$ is the set of all actions' distributions which are
uniformly bounded away from zero by $\delta$.

\begin{corollary}
\label{cor:eps-bne}Let $G=\left(  \left(  T,\lambda\right)  ,A,u,\mathcal{Q}%
,D\right)  $ be a nonatomic\ game with model misspecification and
$\varepsilon>0$.\ If $\lambda$ is strongly continuous, $D$ is strictly convex
in the second argument, and there exists $\delta>0$ such that $Q_{t}%
\subseteq\Delta_{\delta}$ for all $t\in T$,\ then $G$ has\ an $\varepsilon$-BNE.
\end{corollary}

\begin{remark}
\label{rmk:eps-BNE}\emph{Four\ observations are in order:}

\begin{enumerate}
\item \emph{Unlike Esponda and Pouzo \cite{EP} original formulation, in our
definition each player's set of actions' distributions }$Q_{t}$%
\emph{\ does\ not depend on the action played. Conceptually, this amounts to
assume that there is perfect statistical feedback. If we were to impose that
each }$Q_{t}$\emph{ was also function of the action, that is\ }$a\mapsto
Q_{t}\left(  a\right)  $\emph{, the feedback function in (\ref{eq:pro-div})
would fail to satisfy property (\ref{eq:wea-gro}).}

\item \emph{In Definition \ref{def:eps-BNE}, we allow each player's
equilibrium belief }$\beta\left(  t\right)  $\emph{ to be possibly outside the
set }$Q_{t}$\emph{. This could be interpreted as allowing for the possibility
that each player fears model misspecification and willingly considers
probability models\ that are outside his posited set }$Q_{t}$\emph{ (see
Cerreia-Vioglio, Hansen, Maccheroni, and Marinacci \cite{CHMM}). At the same
time, we could have considered the following more stringent definition of
}$\varepsilon$\emph{-BNE where this is not allowed. In this case, }$\sigma
\in\Sigma$\emph{ would be an }$\varepsilon$\emph{-BNE if and only if there
exists a profile of beliefs }$\beta\in\Delta^{T}$\emph{ satisfying}%
\[
\lambda\left(  \left\{  t\in T\ \left\vert \
\begin{array}
[c]{l}%
u_{t}\left(  \sigma\left(  t\right)  ,\beta\left(  t\right)  \right)  \geq
u_{t}\left(  a,\beta\left(  t\right)  \right)  \quad\forall a\in A\\
d_{\Delta}\left(  \beta\left(  t\right)  ,\operatorname{argmin}_{z\in Q_{t}%
}D\left(  \lambda_{\sigma}||z\right)  \right)  \leq\varepsilon\text{\emph{ and
}}\beta\left(  t\right)  \in Q_{t}%
\end{array}
\right.  \right\}  \right)  =1
\]
\emph{Under the same exact assumptions of Corollary \ref{cor:eps-bne}, we can
show that also these }$\varepsilon$\emph{-equilibria exist.}

\item \emph{Our results do not directly apply to the case in which }$D$\emph{
is the Kullback--Leibler distance }$K$\emph{. In fact, in this case,
}$K\left(  x||\cdot\right)  $\emph{ might fail to be strictly
convex.\footnote{In fact, $K\left(  x||\cdot\right)  $ is strictly convex if
$x\in\Delta^{o}$, but it might fail to be so if $x\in\Delta\backslash
\Delta^{o}$.} At the same time, any perturbation }$\kappa>0\emph{\ }$\emph{of
a statistical divergence }$D$\emph{, that is }$D+\kappa d_{\Delta}^{2}$\emph{,
is a statistical divergence\ and satisfies\ the condition of strict convexity
in Corollary \ref{cor:eps-bne}.}

\item \emph{The assumption \textquotedblleft there exists }$\delta>0$\emph{
such that }$Q_{t}\subseteq\Delta_{\delta}$\emph{ for all }$t\in T$%
\emph{\textquotedblright\ is equivalent to the condition \textquotedblleft
each }$Q$\emph{ that belongs to the Hausdorff distance closure of
}$\mathcal{Q}$\emph{ is a subset of }$\Delta^{o}$\emph{\textquotedblright. In
other words, it is an assumption of relative compactness.}\hfill
$\blacktriangle$
\end{enumerate}
\end{remark}

\appendix

\section{Appendix}

In what follows, we first provide the proofs of the results in the main text
and then conclude with one\ of the authors explaining the origin of nonatomic
games. We begin with Appendix \ref{app:bel} where we discuss a result which is
key in proving Theorem \ref{thm:mai}. Appendix \ref{app:pro} contains the
remaining proofs. In a nutshell, this latter section is divided into\ two
parts. First, we deal with the proof of existence of $\varepsilon$-estimated
equilibria. Second, we prove the existence of $\varepsilon$-SCE, $\varepsilon
$-NE, $\varepsilon$-PCE, and $\varepsilon$-BNE by showing that they are all
particular cases or consequence of the existence of $\varepsilon$-estimated equilibria.

In the appendix, the vector space we use is the Cartesian product of $m$
copies of $%
\mathbb{R}
^{n}$, that is $\left(
\mathbb{R}
^{n}\right)  ^{m}$, where $n$\ is given by the cardinality of the space of
actions $A$\ and $m$ is given by the cardinality of the
neighborhood\ structure $\left(  \Pi,\pi\right)  $. We denote the elements of
$\left(
\mathbb{R}
^{n}\right)  ^{m}$ by bold letters, that is $\mathbf{x}$ and $\mathbf{y}$,
while $x_{j}$ will be the vector in $%
\mathbb{R}
^{n}$ which is the $j$-th component of $\mathbf{x}$.\ If $m=1$, then we
denote\ $\mathbf{x}$ and $\mathbf{y}$ simply by $x$ and $y$. We endow $\left(
%
\mathbb{R}
^{n}\right)  ^{m}$\ with the topology induced by the norm $\left\Vert
\mathbf{x}\right\Vert =\max_{j\in\left\{  1,...,m\right\}  }\left\Vert
x_{j}\right\Vert _{2}$ where $\left\Vert \text{ }\right\Vert _{2}$ is the
Euclidean norm. Finally, we denote the Cartesian product of $m$ copies of
$\Delta$ by $\Delta^{m}$.\ Note that $\Delta^{m}$ is a nonempty, convex, and
compact subset of $\left(
\mathbb{R}
^{n}\right)  ^{m}$ and we endow it with the distance induced by $\left\Vert
\text{ }\right\Vert $.

\subsection{A key general result\label{app:bel}}

The next lemma uses the terminology of Bhaskara Rao and Bhaskara Rao
\cite{RR}. Before discussing it, we need a piece of notation which will turn
out to be useful in our later analysis. If $T$ and $A$ are two generic
nonempty sets and $\Gamma:T\rightrightarrows A$ is a (nonempty valued)
correspondence, we denote by $\mathrm{Sel}\left(  \Gamma\right)  $ the set of
all selections of $\Gamma$, that is, the set of all functions $\gamma
:T\rightarrow A$ such that $\gamma\left(  t\right)  \in\Gamma\left(  t\right)
$ for all $t\in T$. Just for this section, $\mathcal{T}$ is an arbitrary
$\sigma$-algebra of subsets\ of $T$.\footnote{In the rest of the paper,
$\mathcal{T}$\ is the power set.} Finally, given a $\mathcal{T}$-measurable
map $\gamma:T\rightarrow A$ and a probability $\mu:\mathcal{T}\rightarrow
\left[  0,1\right]  $, recall that%
\[
\mu_{\gamma}=\left(  \mu\left(  \left\{  t\in T\mid\gamma(t)=a\right\}
\right)  \right)  _{a\in A}%
\]

\begin{lemma}
\label{lem:believe-me}Let $\left(  T,\mathcal{T}\right)  $ be a measurable
space, $A$\ a finite set with $n$ elements, and $\boldsymbol{\lambda}=\left(
\lambda^{1},...,\lambda^{m}\right)  $ a vector of strongly continuous
probabilities on $\mathcal{T}$. If $\Gamma:T\rightrightarrows A$ is a
correspondence, then%
\[
\left\{  \boldsymbol{\lambda}_{\gamma}=\left(  \lambda_{\gamma}^{1}%
,...,\lambda_{\gamma}^{m}\right)  \mid\gamma\in\mathrm{Sel}\left(
\Gamma\right)  \text{ and }\gamma\text{ is }\mathcal{T}\text{-measurable}%
\right\}
\]
is a convex subset of $\Delta^{m}$.
\end{lemma}

\noindent\textbf{Proof.} If $\phi,\gamma\in\mathrm{Sel}\left(  \Gamma\right)
$ and are $\mathcal{T}$-measurable, for each $\alpha\in\left(  0,1\right)  $,
we want to construct $\psi\in\mathrm{Sel}\left(  \Gamma\right)  $ which is
$\mathcal{T}$-measurable and such that $\boldsymbol{\lambda}_{\psi}%
=\alpha\boldsymbol{\lambda}_{\phi}+\left(  1-\alpha\right)
\boldsymbol{\lambda}_{\gamma}$.

Set $S_{ij}=\phi^{-1}\left(  i\right)  \cap\gamma^{-1}\left(  j\right)  $ for
all $i,j\in A$. Then $\left\{  S_{ij}\right\}  _{i,j\in A}$ forms a partition
of $T$ (with possibly some empty elements) and all its elements belong to
$\mathcal{T}$, because $\phi^{-1}\left(  i\right)  ,\gamma^{-1}\left(
j\right)  \in\mathcal{T}$ for all $i,j\in A$.

Since $\lambda^{1},...,\lambda^{m}$ are strongly continuous and $\mathcal{T}$
is a $\sigma$-algebra, for any $i,j\in A$, there are $T_{ij},U_{ij}%
\in\mathcal{T}$ such that $S_{ij}=T_{ij}\sqcup U_{ij}$,\footnote{$\sqcup$
denotes the disjoint union.} $\boldsymbol{\lambda}\left(  T_{ij}\right)
=\alpha\boldsymbol{\lambda}\left(  S_{ij}\right)  $, and $\boldsymbol{\lambda
}\left(  U_{ij}\right)  =\left(  1-\alpha\right)  \boldsymbol{\lambda}\left(
S_{ij}\right)  $. This is trivial if $S_{ij}$ is empty, else set
\begin{align*}
\mathcal{T}_{ij}  &  =\mathcal{T}\cap S_{ij}\\
\lambda_{ij}^{k}\left(  S\right)   &  =\lambda^{k}\left(  S\right)
\qquad\forall S\in\mathcal{T}_{ij},\forall k=1,...,m
\end{align*}
and notice that $\lambda_{ij}^{1},...,\lambda_{ij}^{m}$ are strongly
continuous, positive, and bounded charges on the $\sigma$-algebra
$\mathcal{T}_{ij}$. By Bhaskara Rao and Bhaskara Rao \cite[Theorem 11.4.9]%
{RR}, the set
\[
R\left(  \boldsymbol{\lambda}_{ij}\right)  =\left\{  \left(  \lambda_{ij}%
^{1}\left(  S\right)  ,...,\lambda_{ij}^{m}\left(  S\right)  \right)  \mid
S\in\mathcal{T}_{ij}\right\}
\]
is convex in $\mathbb{R}^{m}$. Moreover, both $\boldsymbol{0}=\left(
\lambda_{ij}^{1}\left(  \varnothing\right)  ,...,\lambda_{ij}^{m}\left(
\varnothing\right)  \right)  $ and $\boldsymbol{\lambda}_{ij}\left(
S_{ij}\right)  =\left(  \lambda_{ij}^{1}\left(  S_{ij}\right)  ,...,\lambda
_{ij}^{m}\left(  S_{ij}\right)  \right)  $ belong to $R\left(
\boldsymbol{\lambda}_{ij}\right)  $. By convexity of the latter, there exists
$T_{ij}\in\mathcal{T}_{ij}$ such that $\boldsymbol{\lambda}_{ij}\left(
T_{ij}\right)  =\alpha\boldsymbol{\lambda}_{ij}\left(  S_{ij}\right)  $. But
then $T_{ij},U_{ij}=S_{ij}\setminus T_{ij}\in\mathcal{T}$, $S_{ij}%
=T_{ij}\sqcup U_{ij}$, $\boldsymbol{\lambda}\left(  T_{ij}\right)
=\boldsymbol{\lambda}_{ij}\left(  T_{ij}\right)  =\alpha\boldsymbol{\lambda
}_{ij}\left(  S_{ij}\right)  =\alpha\boldsymbol{\lambda}\left(  S_{ij}\right)
$, and $\boldsymbol{\lambda}\left(  U_{ij}\right)  =\left(  1-\alpha\right)
\boldsymbol{\lambda}\left(  S_{ij}\right)  $ by additivity of
$\boldsymbol{\lambda}$.

The function\ $\psi:T\rightarrow A$ defined by%
\[
\psi\left(  t\right)  =\left\{
\begin{array}
[c]{c}%
\phi\left(  t\right)  =i\quad\text{if }t\in T_{ij}\\
\gamma\left(  t\right)  =j\quad\text{if }t\in U_{ij}%
\end{array}
\right.
\]
is well defined and $\psi\left(  t\right)  \in\left\{  \phi\left(  t\right)
,\gamma\left(  t\right)  \right\}  \subseteq\Gamma\left(  t\right)  $ for all
$t\in T$, so that $\psi\in\mathrm{Sel}\left(  \Gamma\right)  $. For each $k\in
A$,%
\begin{align*}
\psi^{-1}\left(  k\right)   &  =\left\{  t\in T\mid\psi\left(  t\right)
=k\right\}  =\left\{  t\in\left(
{\displaystyle\bigsqcup\limits_{i,j\in A}}
T_{ij}\right)  \sqcup\left(
{\displaystyle\bigsqcup\limits_{i,j\in A}}
U_{ij}\right)  \mid\psi\left(  t\right)  =k\right\} \\
&  =\left(
{\displaystyle\bigsqcup\limits_{i,j\in A}}
\left\{  t\in T_{ij}\mid\psi\left(  t\right)  =k\right\}  \right)
\sqcup\left(
{\displaystyle\bigsqcup\limits_{i,j\in A}}
\left\{  t\in U_{ij}\mid\psi\left(  t\right)  =k\right\}  \right) \\
&  =\left(
{\displaystyle\bigsqcup\limits_{i,j\in A}}
\left\{  t\in T_{ij}\mid\phi\left(  t\right)  =k\right\}  \right)
\sqcup\left(
{\displaystyle\bigsqcup\limits_{i,j\in A}}
\left\{  t\in U_{ij}\mid\gamma\left(  t\right)  =k\right\}  \right)
\end{align*}
but, for all $t\in T_{ij}$, $\phi\left(  t\right)  =i$, then

\begin{itemize}
\item if $i=k$, $\left\{  t\in T_{ij}\mid\phi\left(  t\right)  =k\right\}
=T_{ij}$,

\item else $i\neq k$ and $\left\{  t\in T_{ij}\mid\phi\left(  t\right)
=k\right\}  =\varnothing$,
\end{itemize}

\noindent thus $%
{\displaystyle\bigsqcup\limits_{i,j\in A}}
\left\{  t\in T_{ij}\mid\phi\left(  t\right)  =k\right\}  =%
{\displaystyle\bigsqcup\limits_{i,j\in A\mid i=k}}
T_{ij}=%
{\displaystyle\bigsqcup\limits_{j\in A}}
T_{kj}$; analogously, for all $t\in U_{ij}$, $\gamma\left(  t\right)  =j$;\ then

\begin{itemize}
\item if $j=k$, $\left\{  t\in U_{ij}\mid\gamma\left(  t\right)  =k\right\}
=U_{ij}$,

\item else $j\neq k$ and $\left\{  t\in U_{ij}\mid\gamma\left(  t\right)
=k\right\}  =\varnothing$,
\end{itemize}

\noindent thus $%
{\displaystyle\bigsqcup\limits_{i,j\in A}}
\left\{  t\in U_{ij}\mid\gamma\left(  t\right)  =k\right\}  =%
{\displaystyle\bigsqcup\limits_{i,j\in A\mid j=k}}
U_{ij}=%
{\displaystyle\bigsqcup\limits_{i\in A}}
U_{ik}$; therefore,%
\[
\psi^{-1}\left(  k\right)  =\left(
{\displaystyle\bigsqcup\limits_{j\in A}}
T_{kj}\right)  \sqcup\left(
{\displaystyle\bigsqcup\limits_{i\in A}}
U_{ik}\right)  \in\mathcal{T}%
\]
As a consequence, $\psi$ is $\mathcal{T}$-measurable and, for each $k\in A$,
and for\ each $l=1,...,m$,%
\begin{align*}
\lambda^{l}\left(  \psi^{-1}\left(  k\right)  \right)   &  =%
{\displaystyle\sum\limits_{j\in A}}
\lambda^{l}\left(  T_{kj}\right)  +%
{\displaystyle\sum\limits_{i\in A}}
\lambda^{l}\left(  U_{ik}\right)  =%
{\displaystyle\sum\limits_{j\in A}}
\alpha\lambda^{l}\left(  S_{kj}\right)  +%
{\displaystyle\sum\limits_{i\in A}}
\left(  1-\alpha\right)  \lambda^{l}\left(  S_{ik}\right) \\
&  =\alpha\lambda^{l}\left(
{\displaystyle\bigsqcup\limits_{j\in A}}
S_{kj}\right)  +\left(  1-\alpha\right)  \lambda^{l}\left(
{\displaystyle\bigsqcup\limits_{i\in A}}
S_{ik}\right) \\
&  =\alpha\lambda^{l}\left(
{\displaystyle\bigsqcup\limits_{j\in A}}
\left(  \phi^{-1}\left(  k\right)  \cap\gamma^{-1}\left(  j\right)  \right)
\right)  +\left(  1-\alpha\right)  \lambda^{l}\left(
{\displaystyle\bigsqcup\limits_{i\in A}}
\left(  \phi^{-1}\left(  i\right)  \cap\gamma^{-1}\left(  k\right)  \right)
\right) \\
&  =\alpha\lambda^{l}\left(  \phi^{-1}\left(  k\right)  \cap%
{\displaystyle\bigsqcup\limits_{j\in A}}
\gamma^{-1}\left(  j\right)  \right)  +\left(  1-\alpha\right)  \lambda
^{l}\left(  \gamma^{-1}\left(  k\right)  \cap%
{\displaystyle\bigsqcup\limits_{i\in A}}
\phi^{-1}\left(  i\right)  \right) \\
&  =\alpha\lambda^{l}\left(  \phi^{-1}\left(  k\right)  \cap T\right)
+\left(  1-\alpha\right)  \lambda^{l}\left(  \gamma^{-1}\left(  k\right)  \cap
T\right) \\
&  =\alpha\lambda^{l}\left(  \phi^{-1}\left(  k\right)  \right)  +\left(
1-\alpha\right)  \lambda^{l}\left(  \gamma^{-1}\left(  k\right)  \right)
\end{align*}
thus $\lambda_{\psi}^{l}=\alpha\lambda_{\phi}^{l}+\left(  1-\alpha\right)
\lambda_{\gamma}^{l}$. Since this is true for each $l=1,...,m$, then
$\boldsymbol{\lambda}_{\psi}=\alpha\boldsymbol{\lambda}_{\phi}+\left(
1-\alpha\right)  \boldsymbol{\lambda}_{\gamma}$, as wanted.\hfill
$\blacksquare$

$\bigskip$

Building on this lemma, Gilboa, Maccheroni, Marinacci, and Schmeidler
\cite{GMMS} prove that, when $m=1$, $\left\{  \lambda_{\gamma}\mid\gamma
\in\mathrm{Sel}\left(  \Gamma\right)  \text{ and }\gamma\text{ is }%
\mathcal{T}\text{-measurable}\right\}  $ is indeed the core of the belief
function%
\[
\operatorname*{Bel}\left(  I\right)  =\lambda\left(  \left\{  t\in T\mid
\Gamma\left(  t\right)  \subseteq I\right\}  \right)  \quad\forall I\subseteq
A
\]
and they characterize its extreme points \emph{\`{a} la} Shapley \cite{Sha}.

\subsection{Proofs and related material\label{app:pro}}

In what follows and up to the proof of Corollary \ref{cor:eps-sce}, we
consider a nonatomic game with estimation feedback $G=\left(  \left(
T,\lambda\right)  ,A,u,\left(  \Pi,\pi\right)  ,f\right)  $.\ Recall that
$\Pi$ is a collection of nonempty subsets of $T$, $\left\{  T_{j}\right\}
_{j=1}^{m}$, such that $\lambda\left(  T_{j}\right)  >0$ for all $j\in\left\{
1,...,m\right\}  $ and $T=\cup_{j=1}^{m}T_{j}$. The proof of existence of
$\varepsilon$-estimated equilibria rests on two key ideas which we formally
develop below:

\begin{enumerate}
\item We first consider different correspondences and study their properties.
This study\ culminates\ with the correspondence $\widetilde{\mathbf{BR}%
}_{f,\varepsilon}:\Delta^{m}\rightrightarrows\Delta^{m}$ defined in
(\ref{tilda br sta}) below.\ All of these correspondences are basically
$\varepsilon$-consistent/confirmed best-reply correspondences. To fix ideas,
for the case $\Pi=\left\{  T\right\}  $ and $m=1$, in words, given $x\in
\Delta$ and $y\in\widetilde{\mathbf{BR}}_{f,\varepsilon}\left(  x\right)  $,
$y$ is a possible\ distribution of strategies\ in the population, which arises
if the players' distribution of actions was $x$ and players best-responded to
it\ using a belief which was $\varepsilon$-consistent\ with respect to $x$.

\item We then show that $\widetilde{\mathbf{BR}}_{f,\varepsilon}$ has a fixed
point by using Browder's\ Fixed\ Point\ Theorem.\ This will give us the
equilibrium in pure strategies that we are after.
\end{enumerate}

Consider $\varepsilon>0$. First, let $\mathbf{BR}_{f,\varepsilon}%
:T\times\Delta^{m}\rightrightarrows A$ be defined by%
\begin{align*}
&  \mathbf{BR}_{f,\varepsilon}\left(  t,\mathbf{x}\right) \\
&  =\left\{  b\in A\mid\exists\beta_{t}\in\Delta\text{ s.t. }f_{t}\left(
b,\beta_{t},x_{\pi\left(  t\right)  }\right)  <\varepsilon\text{\ and }%
u_{t}\left(  b,\beta_{t}\right)  \geq u_{t}\left(  a,\beta_{t}\right)
\quad\forall a\in A\right\}
\end{align*}
for all $\left(  t,\mathbf{x}\right)  \in T\times\Delta^{m}$. Clearly,
$\mathbf{BR}_{f,\varepsilon}\left(  t,\mathbf{x}\right)  $ is the set of all
pure strategies which are a best-reply of player $t$\ to some belief\ $\beta
_{t}$ where $\beta_{t}$\ is $\varepsilon$-consistent when assuming the true
distribution restricted to the subpopulation $T_{\pi\left(  t\right)  }$\ is
$x_{\pi\left(  t\right)  }$. One can derive several related \textquotedblleft%
$\varepsilon$-consistent best-reply\textquotedblright\ correspondences from
this basic one. For each $\mathbf{x}\in\Delta^{m}$, denote the $\mathbf{x}%
$-section $\mathbf{BR}_{f,\varepsilon}\left(  \cdot,\mathbf{x}\right)
:T\rightrightarrows A$\ of $\mathbf{BR}_{f,\varepsilon}$ by $\mathbf{BR}%
_{f,\varepsilon}^{\mathbf{x}}$. Next, let $\Phi_{f,\varepsilon}:\Delta
^{m}\rightrightarrows\Sigma$ be defined as $\Phi_{f,\varepsilon}%
(\mathbf{x})=\mathrm{Sel}\left(  \mathbf{BR}_{f,\varepsilon}^{\mathbf{x}%
}\right)  $ for all $\mathbf{x}\in\Delta^{m}$ where $\mathrm{Sel}\left(
\mathbf{BR}_{f,\varepsilon}^{\mathbf{x}}\right)  $\ is the set of all
selections of $\mathbf{BR}_{f,\varepsilon}^{\mathbf{x}}$. Thus, for a strategy
profile $\sigma\in\Sigma$, we have that%
\begin{align*}
&  \left[  \sigma\in\Phi_{f,\varepsilon}\left(  \mathbf{x}\right)  \right] \\
&  \iff\left[  \forall t\in T,\ \sigma(t)\in\mathbf{BR}_{f,\varepsilon
}^{\mathbf{x}}\left(  t\right)  \right] \\
&  \iff\left[  \forall t\in T,\ \sigma(t)\in\mathbf{BR}_{f,\varepsilon}\left(
t,\mathbf{x}\right)  \right] \\
&  \iff\left[  \forall t\in T\ \exists\beta_{t}\in\Delta\text{ s.t. }%
f_{t}\left(  \sigma\left(  t\right)  ,\beta_{t},x_{\pi\left(  t\right)
}\right)  <\varepsilon\ \text{and }u_{t}\left(  \sigma(t),\beta_{t}\right)
\geq u_{t}\left(  a,\beta_{t}\right)  \quad\forall a\in A\right]
\end{align*}

\begin{remark}
\label{ann:oso-sta}\emph{If there exists }$\mathbf{x}\in\Delta^{m}%
$\ \emph{such that\ }$\sigma\in\Phi_{f,\varepsilon}\left(  \mathbf{x}\right)
$\emph{ and }$x_{\pi\left(  t\right)  }=\lambda_{\sigma}^{\pi\left(  t\right)
}$\emph{ for all }$t\in T$\emph{, then }$\sigma$\emph{ is an }$\varepsilon
$\emph{-estimated equilibrium. In fact, we have}

\begin{enumerate}
\item \emph{For each }$t\in T$\emph{ there exists }$\beta_{t}\in\Delta$\emph{
such that }$f_{t}\left(  \sigma\left(  t\right)  ,\beta_{t},x_{\pi\left(
t\right)  }\right)  <\varepsilon\ $\emph{and }$u_{t}\left(  \sigma
(t),\beta_{t}\right)  \geq u_{t}\left(  a,\beta_{t}\right)  $\emph{ for
all}$\ a\in A$\emph{;}

\item \emph{We can define }$\beta:T\rightarrow\Delta$\emph{ by }$\beta\left(
t\right)  =\beta_{t}$\emph{ for all }$t\in T$\emph{.}
\end{enumerate}

\noindent\emph{This implies that for each }$t\in T$

\begin{enumerate}
\item[a)] $u_{t}\left(  \sigma(t),\beta\left(  t\right)  \right)  \geq
u_{t}\left(  a,\beta\left(  t\right)  \right)  $\emph{ for all}$\ a\in
A$\emph{ (optimality);}

\item[b)] $f_{t}\left(  \sigma\left(  t\right)  ,\beta\left(  t\right)
,\lambda_{\sigma}^{\pi\left(  t\right)  }\right)  <\varepsilon$\emph{ (strict
}$\varepsilon$\emph{-consistency),}
\end{enumerate}

\noindent\emph{that is, }$\left\{  t\in T\ \left\vert \
\begin{array}
[c]{l}%
u_{t}\left(  \sigma\left(  t\right)  ,\beta\left(  t\right)  \right)  \geq
u_{t}\left(  a,\beta\left(  t\right)  \right)  \quad\forall a\in A\\
f_{t}\left(  \sigma\left(  t\right)  ,\beta\left(  t\right)  ,\lambda_{\sigma
}^{\pi\left(  t\right)  }\right)  <\varepsilon
\end{array}
\right.  \right\}  =T$\emph{. In particular, it holds that}%
\[
\lambda\left(  \left\{  t\in T\ \left\vert \
\begin{array}
[c]{l}%
u_{t}\left(  \sigma\left(  t\right)  ,\beta\left(  t\right)  \right)  \geq
u_{t}\left(  a,\beta\left(  t\right)  \right)  \quad\forall a\in A\\
f_{t}\left(  \sigma\left(  t\right)  ,\beta\left(  t\right)  ,\lambda_{\sigma
}^{\pi\left(  t\right)  }\right)  \leq\varepsilon
\end{array}
\right.  \right\}  \right)  =1
\]
\hfill$\blacktriangle$
\end{remark}

Next, consider the correspondence $\mathbf{B}_{f,\varepsilon}:\Delta
^{m}\rightrightarrows\Sigma$ defined by%
\[
\mathbf{B}_{f,\varepsilon}\left(  \mathbf{x}\right)  =\left\{  \sigma\in
\Sigma\mid\exists\beta\in\Delta^{T}\text{\ s.t. }%
\begin{array}
[c]{l}%
u_{t}\left(  \sigma(t),\beta\left(  t\right)  \right)  \geq u_{t}\left(
a,\beta\left(  t\right)  \right)  \quad\forall a\in A,\forall t\in T\\
\sup_{t\in T}f_{t}\left(  \sigma\left(  t\right)  ,\beta\left(  t\right)
,x_{\pi\left(  t\right)  }\right)  <\varepsilon
\end{array}
\right\}  \quad\forall\mathbf{x}\in\Delta^{m}%
\]

\begin{lemma}
\label{lem:B-dec-sta}$\mathbf{B}_{f,\varepsilon}\left(  \mathbf{x}\right)
=\bigcup_{\eta\in\left(  0,\varepsilon\right)  }\Phi_{f,\eta}\left(
\mathbf{x}\right)  \subseteq\Phi_{f,\varepsilon}\left(  \mathbf{x}\right)
$\ for all $\mathbf{x}\in\Delta^{m}$.
\end{lemma}

\noindent\textbf{Proof.} Fix $\mathbf{x}\in\Delta^{m}$. Consider $\sigma
\in\bigcup_{\eta\in\left(  0,\varepsilon\right)  }\Phi_{f,\eta}\left(
\mathbf{x}\right)  $. It follows that $\sigma\in\Phi_{f,\eta}\left(
\mathbf{x}\right)  $ for some $\eta\in\left(  0,\varepsilon\right)  $. This
implies that $\sigma\in\Sigma$\ and $\sigma\left(  t\right)  \in
\mathbf{BR}_{f,\eta}(t,\mathbf{x})$ for all\ $t\in T$, that is, for each $t\in
T$ there exists $\beta_{t}\in\Delta$\ such that\ $f_{t}\left(  \sigma\left(
t\right)  ,\beta_{t},x_{\pi\left(  t\right)  }\right)  <\eta\ $and
$u_{t}\left(  \sigma(t),\beta_{t}\right)  \geq u_{t}\left(  a,\beta
_{t}\right)  $ for all $a\in A$. In particular, if we define $\beta\in
\Delta^{T}$ as $\beta\left(  t\right)  =\beta_{t}$ for all $t\in T$, we have
that $u_{t}\left(  \sigma(t),\beta\left(  t\right)  \right)  \geq u_{t}\left(
a,\beta\left(  t\right)  \right)  $ for all $a\in A$ and for all $t\in T$, and%
\[
\sup_{t\in T}f_{t}\left(  \sigma\left(  t\right)  ,\beta\left(  t\right)
,x_{\pi\left(  t\right)  }\right)  \leq\eta<\varepsilon
\]
yielding\ that $\sigma\in\mathbf{B}_{f,\varepsilon}\left(  \mathbf{x}\right)
$. Conversely, if $\sigma\in\mathbf{B}_{f,\varepsilon}\left(  \mathbf{x}%
\right)  $, then there exists $\beta\in\Delta^{T}$\ such that%
\begin{equation}
\sup_{t\in T}f_{t}\left(  \sigma\left(  t\right)  ,\beta\left(  t\right)
,x_{\pi\left(  t\right)  }\right)  <\varepsilon\label{eq:sup}%
\end{equation}
and $u_{t}\left(  \sigma(t),\beta\left(  t\right)  \right)  \geq u_{t}\left(
a,\beta\left(  t\right)  \right)  $ for all $a\in A$ and for all $t\in T$. It
follows that there exists $\bar{\eta}\in\left(  0,\varepsilon\right)  $ such
that (\ref{eq:sup}) holds with $\bar{\eta}$\ in place of $\varepsilon$. This
implies that $\sigma\in\Phi_{f,\bar{\eta}}\left(  \mathbf{x}\right)
\subseteq\bigcup_{\eta\in\left(  0,\varepsilon\right)  }\Phi_{f,\eta}\left(
\mathbf{x}\right)  $.

Obviously, if $0<\eta<\eta^{\prime}$, then $\mathbf{BR}_{f,\eta}\left(
t,\mathbf{x}\right)  \subseteq\mathbf{BR}_{f,\eta^{\prime}}\left(
t,\mathbf{x}\right)  $ for all $t\in T$ and for all $\mathbf{x}\in\Delta^{m}$
and, in particular, $\Phi_{f,\eta}\left(  \mathbf{x}\right)  \subseteq
\Phi_{f,\eta^{\prime}}\left(  \mathbf{x}\right)  $.\ This implies that
$\bigcup_{\eta\in\left(  0,\varepsilon\right)  }\Phi_{f,\eta}\left(
\mathbf{x}\right)  \subseteq\Phi_{f,\varepsilon}\left(  \mathbf{x}\right)
$.\hfill$\blacksquare$

\bigskip

Remark\ \ref{ann:oso-sta}\ above will be useful to justify the following last
correspondence: $\widetilde{\mathbf{BR}}_{f,\varepsilon}:\Delta^{m}%
\rightrightarrows\Delta^{m}$ defined by%
\begin{equation}
\widetilde{\mathbf{BR}}_{f,\varepsilon}\left(  \mathbf{x}\right)  =\left\{
\mathbf{y}\in\Delta^{m}\mid\exists\sigma\in\mathbf{B}_{f,\varepsilon}\left(
\mathbf{x}\right)  \text{ s.t.\ }\lambda_{\sigma}^{j}=y_{j}\quad\forall
j\in\left\{  1,...,m\right\}  \right\}  \quad\forall\mathbf{x}\in\Delta^{m}
\label{tilda br sta}%
\end{equation}
In other words, $\widetilde{\mathbf{BR}}_{f,\varepsilon}\left(  \mathbf{x}%
\right)  $ is the collection of actions' distributions $\mathbf{y}=\left(
y_{j}\right)  _{j=1}^{m}$\ on the subpopulations of players, which can be
induced by the $\beta$-optimal choice of strategies $\sigma$ where beliefs
$\beta=\left(  \beta_{t}\right)  _{t\in T}$ are close enough in terms of
feedback to $\mathbf{x}=\left(  x_{j}\right)  _{j=1}^{m}$. Note that%
\begin{align}
\widetilde{\mathbf{BR}}_{f,\varepsilon}\left(  \mathbf{x}\right)   &
=\left\{  \left(  \lambda_{\sigma}^{j}\right)  _{j=1}^{m}\mid\sigma
\in\mathbf{B}_{f,\varepsilon}\left(  \mathbf{x}\right)  \right\}  =\left\{
\left(  \lambda_{\sigma}^{j}\right)  _{j=1}^{m}\mid\sigma\in\bigcup_{\eta
\in\left(  0,\varepsilon\right)  }\Phi_{f,\eta}\left(  \mathbf{x}\right)
\right\} \label{eq:til-br-cha-sta}\\
&  =\bigcup_{\eta\in\left(  0,\varepsilon\right)  }\left\{  \left(
\lambda_{\sigma}^{j}\right)  _{j=1}^{m}\mid\sigma\in\Phi_{f,\eta}\left(
\mathbf{x}\right)  \right\}  =\bigcup_{\eta\in\left(  0,\varepsilon\right)
}\left\{  \left(  \lambda_{\sigma}^{j}\right)  _{j=1}^{m}\mid\sigma
\in\mathrm{Sel}\left(  \mathbf{BR}_{f,\eta}^{\mathbf{x}}\right)  \right\}
\label{eq:union-jack-sta}%
\end{align}
An immediate implication of the definition in (\ref{tilda br sta}) is the next result.

\begin{lemma}
\label{lem:BR-equ-sta}If $\mathbf{x}\in\widetilde{\mathbf{BR}}_{f,\varepsilon
}(\mathbf{x})$, then there exists\ an\ $\varepsilon$-estimated equilibrium
$\sigma$\ such that $\lambda_{\sigma}^{j}=x_{j}$ for all $j\in\left\{
1,..,m\right\}  $.
\end{lemma}

\noindent\textbf{Proof. }By Lemma \ref{lem:B-dec-sta} and the definition of
$\widetilde{\mathbf{BR}}_{f,\varepsilon}$, if $\mathbf{x}\in
\widetilde{\mathbf{BR}}_{f,\varepsilon}(\mathbf{x})$, then there exists
$\sigma\in\mathbf{B}_{f,\varepsilon}\left(  \mathbf{x}\right)  \subseteq
\Phi_{f,\varepsilon}\left(  \mathbf{x}\right)  $ such that\ $\lambda_{\sigma
}^{j}=x_{j}$ for all $j\in\left\{  1,...,m\right\}  $%
.\ Remark\ \ref{ann:oso-sta} yields\ that $\sigma$ is an $\varepsilon
$-estimated equilibrium.\hfill$\blacksquare$

\begin{lemma}
\label{lem:non-emp-cvx-sta}If $\lambda$ is strongly continuous, then
$\widetilde{\mathbf{BR}}_{f,\varepsilon}(\mathbf{x})$ is nonempty and convex
for all $\mathbf{x}\in\Delta^{m}$.
\end{lemma}

\noindent\textbf{Proof.} Fix $\mathbf{x}\in\Delta^{m}$\ and $\eta\in\left(
0,\varepsilon\right)  $. Since $f$ satisfies (\ref{eq:wea-gro}), recall that%
\begin{equation}
\forall t\in T,\forall z\in\Delta,\exists\gamma_{t,z}\in\Delta\text{ s.t.
}\forall a\in A\text{ }f_{t}\left(  a,\gamma_{t,z},z\right)  =0
\label{eq:wea-pro-app}%
\end{equation}
Since $\mathbf{x}$\ is given, define $\beta\in\Delta^{T}$ to be such that
$\beta\left(  t\right)  =\gamma_{t,x_{\pi\left(  t\right)  }}$ for all $t\in
T$. Note that\ $\beta\left(  t\right)  \in\Delta$ satisfies $f_{t}\left(
a,\beta\left(  t\right)  ,x_{\pi\left(  t\right)  }\right)  =0<\eta$ for all
$a\in A$ and for all $t\in T$. Since $A$\ is finite, for each $t\in T$ choose
$\sigma\left(  t\right)  \in A\ $such that $u_{t}\left(  \sigma\left(
t\right)  ,\beta\left(  t\right)  \right)  \geq u_{t}\left(  a,\beta\left(
t\right)  \right)  $ for all $a\in A$. This defines a function $\sigma
:T\rightarrow A$, that is $\sigma\in\Sigma$, such that $\sigma\in\Phi_{f,\eta
}\left(  \mathbf{x}\right)  $. By Lemma \ref{lem:B-dec-sta}, we have that
$\sigma\in\mathbf{B}_{f,\varepsilon}\left(  \mathbf{x}\right)  $ and $\left(
\lambda_{\sigma}^{j}\right)  _{j=1}^{m}\in\widetilde{\mathbf{BR}%
}_{f,\varepsilon}\left(  \mathbf{x}\right)  $. Convexity is a consequence of
the following two observations:

\begin{enumerate}
\item By Lemma \ref{lem:believe-me} and since each $\lambda^{j}$ is strongly
continuous, recall that $\{\left(  \lambda_{\sigma}^{j}\right)  _{j=1}^{m}%
\mid\sigma\in\mathrm{Sel}\left(  \mathbf{BR}_{f,\eta}^{\mathbf{x}}\right)  \}$
is a convex subset of $\Delta^{m}$ for all $\eta\in\left(  0,\varepsilon
\right)  $.

\item By (\ref{eq:union-jack-sta}), we have that%
\[
\widetilde{\mathbf{BR}}_{f,\varepsilon}\left(  \mathbf{x}\right)
=\bigcup_{\eta\in\left(  0,\varepsilon\right)  }\left\{  \left(
\lambda_{\sigma}^{j}\right)  _{j=1}^{m}\mid\sigma\in\mathrm{Sel}\left(
\mathbf{BR}_{f,\eta}^{\mathbf{x}}\right)  \right\}
\]

\end{enumerate}

It follows that $\widetilde{\mathbf{BR}}_{f,\varepsilon}\left(  \mathbf{x}%
\right)  $ is the union of a chain of convex sets,\footnote{Recall that if
$0<\eta<\eta^{\prime}$, then%
\[
\mathbf{BR}_{f,\eta}\left(  t,\mathbf{x}\right)  \subseteq\mathbf{BR}%
_{f,\eta^{\prime}}\left(  t,\mathbf{x}\right)  \qquad\forall t\in
T,\forall\mathbf{x}\in\Delta^{m}%
\]
This implies that $\mathrm{Sel}\left(  \mathbf{BR}_{f,\eta}^{\mathbf{x}%
}\right)  =\Phi_{f,\eta}\left(  \mathbf{x}\right)  \subseteq\Phi
_{f,\eta^{\prime}}\left(  \mathbf{x}\right)  =\mathrm{Sel}\left(
\mathbf{BR}_{f,\eta^{\prime}}^{\mathbf{x}}\right)  $ for all $\mathbf{x}%
\in\Delta^{m}$.} proving convexity.\hfill$\blacksquare$\bigskip

For the next result recall that\ a)\ $d_{\Delta}$ is the distance on $\Delta$
induced by the Euclidean norm; b) we say that $f=\left(  f_{t}\right)  _{t\in
T}$ is a family of functions which\ is equicontinuous with respect to the
third argument if\ and only if\ for each $\varepsilon>0$ there exists
$\delta_{\varepsilon}>0$ such that%
\[
d_{\Delta}\left(  x,y\right)  <\delta_{\varepsilon}\implies\left\vert
f_{t}\left(  a,\gamma,x\right)  -f_{t}\left(  a,\gamma,y\right)  \right\vert
<\varepsilon\quad\forall t\in T,\forall a\in A,\forall\gamma\in\Delta
\]
The intuition behind the proof of the\ next lemma is that if a strategy
$\sigma$ was $\beta$-optimal and $\beta$ was $\varepsilon$-consistent, given
$\mathbf{x}$, small perturbations of $\mathbf{x}$ do not disrupt optimality
and $\varepsilon$-consistency.

\begin{lemma}
\label{lem:ope-sta}If\ $f=\left(  f_{t}\right)  _{t\in T}\ $is a family of
functions which is equicontinuous with respect to the third argument,\ then
$\widetilde{\mathbf{BR}}_{f,\varepsilon}^{-1}\left(  \mathbf{y}\right)  $\ is
open for all $\mathbf{y}\in\Delta^{m}$.
\end{lemma}

\noindent\textbf{Proof. }Fix $\mathbf{y}\in\Delta^{m}$. Recall that
$\widetilde{\mathbf{BR}}_{f,\varepsilon}^{-1}\left(  \mathbf{y}\right)
=\left\{  \mathbf{x}\in\Delta^{m}\mid\mathbf{y}\in\widetilde{\mathbf{BR}%
}_{f,\varepsilon}\left(  \mathbf{x}\right)  \right\}  $. Note that%
\[
\mathbf{x}\in\widetilde{\mathbf{BR}}_{f,\varepsilon}^{-1}\left(
\mathbf{y}\right)  \iff\mathbf{y}\in\widetilde{\mathbf{BR}}_{f,\varepsilon
}\left(  \mathbf{x}\right)
\]
and $\widetilde{\mathbf{BR}}_{f,\varepsilon}^{-1}\left(  \mathbf{y}\right)  $
is open if and only if \textquotedblleft for each $\mathbf{\bar{x}}$\ such
that $\mathbf{y}\in\widetilde{\mathbf{BR}}_{f,\varepsilon}\left(
\mathbf{\bar{x}}\right)  $, there exists a ball in $\Delta^{m}$ of radius
$\delta$ and center $\mathbf{\bar{x}}$ such that $\mathbf{y}\in
\widetilde{\mathbf{BR}}_{f,\varepsilon}\left(  \mathbf{x}\right)  $%
\emph{\ }for all $\mathbf{x}\in B_{\delta}\left(  \mathbf{\bar{x}}\right)
$\textquotedblright.

Now arbitrarily choose $\mathbf{\bar{x}}$ such that $\mathbf{y}\in
\widetilde{\mathbf{BR}}_{f,\varepsilon}(\mathbf{\bar{x}})$. By definition of
$\widetilde{\mathbf{BR}}_{f,\varepsilon}(\mathbf{\bar{x}})$, there exist
$\sigma\in\mathbf{B}_{f,\varepsilon}\left(  \mathbf{\bar{x}}\right)
\subseteq\Sigma\ $and $\beta\in\Delta^{T}$\ such that

\begin{enumerate}
\item $\lambda_{\sigma}^{j}=y_{j}$ for all $j\in\left\{  1,...,m\right\}  $;

\item $\sup_{t\in T}f_{t}\left(  \sigma\left(  t\right)  ,\beta\left(
t\right)  ,\bar{x}_{\pi\left(  t\right)  }\right)  <\varepsilon$;

\item $u_{t}\left(  \sigma(t),\beta\left(  t\right)  \right)  \geq
u_{t}\left(  a,\beta\left(  t\right)  \right)  $ for all $a\in A$ and for all
$t\in T$.
\end{enumerate}

By point 2, there exists $\bar{\varepsilon}\in\left(  0,\varepsilon\right)  $
such that%
\[
\sup_{t\in T}f_{t}\left(  \sigma\left(  t\right)  ,\beta\left(  t\right)
,\bar{x}_{\pi\left(  t\right)  }\right)  <\bar{\varepsilon}<\varepsilon
\]
Let $\hat{\varepsilon}\in\left(  0,\frac{\varepsilon-\bar{\varepsilon}}%
{2}\right)  $. Since $f=\left(  f_{t}\right)  _{t\in T}$\ is a family of
functions which is equicontinuous with respect to the third argument, there
exists $\delta_{\hat{\varepsilon}}>0$\ such that%
\[
d_{\Delta}\left(  x,y\right)  <\delta_{\hat{\varepsilon}}\implies\left\vert
f_{t}\left(  a,\gamma,x\right)  -f_{t}\left(  a,\gamma,y\right)  \right\vert
<\hat{\varepsilon}\qquad\forall t\in T,\forall a\in A,\forall\gamma\in\Delta
\]
For each $\mathbf{x}\in B_{\delta_{\hat{\varepsilon}}}\left(  \mathbf{\bar{x}%
}\right)  $ note that $d_{\Delta}\left(  x_{j},\bar{x}_{j}\right)
<\delta_{\hat{\varepsilon}}$ for all $j\in\left\{  1,...,m\right\}  $. This
implies that for each $t\in T$\ and for each $\mathbf{x}\in B_{\delta
_{\hat{\varepsilon}}}\left(  \mathbf{\bar{x}}\right)  $%
\[
\left\vert f_{t}\left(  \sigma\left(  t\right)  ,\beta\left(  t\right)
,x_{\pi\left(  t\right)  }\right)  -f_{t}\left(  \sigma\left(  t\right)
,\beta\left(  t\right)  ,\bar{x}_{\pi\left(  t\right)  }\right)  \right\vert
<\hat{\varepsilon}%
\]
Since $f_{t}\geq0$ for all $t\in T$, it follows that for each $t\in T$ and for
each $\mathbf{x}\in B_{\delta_{\hat{\varepsilon}}}\left(  \mathbf{\bar{x}%
}\right)  $%
\begin{align*}
f_{t}\left(  \sigma\left(  t\right)  ,\beta\left(  t\right)  ,x_{\pi\left(
t\right)  }\right)   &  =\left\vert f_{t}\left(  \sigma\left(  t\right)
,\beta\left(  t\right)  ,x_{\pi\left(  t\right)  }\right)  \right\vert \\
&  \leq\left\vert f_{t}\left(  \sigma\left(  t\right)  ,\beta\left(  t\right)
,\bar{x}_{\pi\left(  t\right)  }\right)  \right\vert +\left\vert f_{t}\left(
\sigma\left(  t\right)  ,\beta\left(  t\right)  ,x_{\pi\left(  t\right)
}\right)  -f_{t}\left(  \sigma\left(  t\right)  ,\beta\left(  t\right)
,\bar{x}_{\pi\left(  t\right)  }\right)  \right\vert \\
&  =f_{t}\left(  \sigma\left(  t\right)  ,\beta\left(  t\right)  ,\bar{x}%
_{\pi\left(  t\right)  }\right)  +\left\vert f_{t}\left(  \sigma\left(
t\right)  ,\beta\left(  t\right)  ,x_{\pi\left(  t\right)  }\right)
-f_{t}\left(  \sigma\left(  t\right)  ,\beta\left(  t\right)  ,\bar{x}%
_{\pi\left(  t\right)  }\right)  \right\vert \\
&  <\bar{\varepsilon}+\hat{\varepsilon}%
\end{align*}
This implies that%
\[
\sup_{t\in T}f_{t}\left(  \sigma\left(  t\right)  ,\beta\left(  t\right)
,x_{\pi\left(  t\right)  }\right)  \leq\bar{\varepsilon}+\hat{\varepsilon
}<\frac{\bar{\varepsilon}+\varepsilon}{2}<\varepsilon\quad\forall\mathbf{x}\in
B_{\delta_{\hat{\varepsilon}}}\left(  \mathbf{\bar{x}}\right)
\]
In other words, for each $\mathbf{x}\in B_{\delta_{\hat{\varepsilon}}}\left(
\mathbf{\bar{x}}\right)  $ we have that $\sigma\in\Sigma\ $is such that the
same\ $\beta\in\Delta^{T}$ of above satisfies points 2 and 3, but with
$\mathbf{x}$ in place of $\mathbf{\bar{x}}$. This yields that\ $\sigma
\in\mathbf{B}_{f,\varepsilon}\left(  \mathbf{x}\right)  $ for all
$\mathbf{x}\in B_{\delta_{\hat{\varepsilon}}}\left(  \mathbf{\bar{x}}\right)
$. Since $\mathbf{y}=\left(  y_{j}\right)  _{j=1}^{m}=\left(  \lambda_{\sigma
}^{j}\right)  _{j=1}^{m}$, we obtain that\ $\mathbf{y}\in
\widetilde{\mathbf{BR}}_{f,\varepsilon}\left(  \mathbf{x}\right)  $ for all
$\mathbf{x}\in B_{\delta_{\hat{\varepsilon}}}\left(  \mathbf{\bar{x}}\right)
$,\ proving the statement.\hfill$\blacksquare$

\bigskip

\noindent\textbf{Proof of Theorem \ref{thm:mai}}. By
Lemma\ \ref{lem:BR-equ-sta}, it is enough to show that $\widetilde{\mathbf{BR}%
}_{f,\varepsilon}:\Delta^{m}\rightrightarrows\Delta^{m}$ has a fixed point.
Clearly, $\Delta^{m}\subseteq\left(
\mathbb{R}
^{n}\right)  ^{m}$ is nonempty, compact, and convex. By
Lemmas\ \ref{lem:non-emp-cvx-sta} and \ref{lem:ope-sta}%
,\ $\widetilde{\mathbf{BR}}_{f,\varepsilon}$ has\ nonempty and convex values
and\ $\widetilde{\mathbf{BR}}_{f,\varepsilon}^{-1}\left(  \mathbf{y}\right)
$\ is open for all $\mathbf{y}\in\Delta^{m}$. By Browder's Fixed Point Theorem
for correspondences (see Theorem 1 of Browder\ \cite{Bro}),
$\widetilde{\mathbf{BR}}_{f,\varepsilon}$ has a fixed point.\hfill
$\blacksquare$

\bigskip

We next prove the remaining results of the main text.

\bigskip

\noindent\textbf{Proof of Corollary \ref{cor:eps-sce}}. It is enough to
observe that a nonatomic game with \textit{message} feedback can be mapped
into a nonatomic game with \textit{estimation }feedback where $f$ is defined
as in (\ref{eq:fed-mes}) and $\Pi=\left\{  T_{1}\right\}  $.\footnote{Thus,
$m=1$, $T_{1}=T$, and $\pi\left(  t\right)  =1$ for all $t\in T$.} With this
identification, an\ $\varepsilon$-estimated equilibrium is a self-confirming
$\varepsilon$-equilibrium. By Theorem \ref{thm:mai}, it is then enough to show
that $f=\left(  f_{t}\right)  _{t\in T}$\ is a family of functions which is
equicontinuous with respect to the third argument. Since $m=\left(
m_{t}\right)  _{t\in T}$\ is a family of functions which is equicontinuous
with respect to the second argument, we have that for each $\varepsilon>0$
there exists $\delta_{\varepsilon}>0$ such that%
\begin{equation}
d_{\Delta}\left(  x,y\right)  <\delta_{\varepsilon}\implies d\left(
m_{t}\left(  a,x\right)  ,m_{t}\left(  a,y\right)  \right)  <\varepsilon
\qquad\forall t\in T,\forall a\in A \label{eq:equ-m}%
\end{equation}
Since for each $t\in T$ we have that $f_{t}\left(  a,x,y\right)  =d\left(
m_{t}\left(  a,x\right)  ,m_{t}\left(  a,y\right)  \right)  \ $for all $a\in
A\ $and for all $x,y\in\Delta$, observe that%
\begin{align*}
\left\vert f_{t}\left(  a,\gamma,x\right)  -f_{t}\left(  a,\gamma,y\right)
\right\vert  &  =\left\vert d\left(  m_{t}\left(  a,\gamma\right)
,m_{t}\left(  a,x\right)  \right)  -d\left(  m_{t}\left(  a,\gamma\right)
,m_{t}\left(  a,y\right)  \right)  \right\vert \\
&  \leq d\left(  m_{t}\left(  a,x\right)  ,m_{t}\left(  a,y\right)  \right)
\quad\forall t\in T,\forall a\in A,\forall x,y,\gamma\in\Delta
\end{align*}
By (\ref{eq:equ-m}), we can conclude that for each $\varepsilon>0$ there
exists $\delta_{\varepsilon}>0$ such that%
\begin{align*}
d_{\Delta}\left(  x,y\right)   &  <\delta_{\varepsilon}\implies\\
\left\vert f_{t}\left(  a,\gamma,x\right)  -f_{t}\left(  a,\gamma,y\right)
\right\vert  &  \leq d\left(  m_{t}\left(  a,x\right)  ,m_{t}\left(
a,y\right)  \right)  <\varepsilon\quad\forall t\in T,\forall a\in
A,\forall\gamma\in\Delta
\end{align*}
proving equicontinuity with respect to the third argument of $f$%
.\hfill$\blacksquare$

\bigskip

\noindent\textbf{Proof of Corollary\ \ref{cor:eps-Nas}}. Consider the
nonatomic game $G=\left(  \left(  T,\lambda\right)  ,A,u\right)  $ and
$\varepsilon>0$.\ Since $u=\left(  u_{t}\right)  _{t\in T}$\ is a family of
functions which is equicontinuous with respect to the second argument, we have
that for each $\hat{\varepsilon}>0$ there exists $\delta_{\hat{\varepsilon}%
}>0$ such that%
\begin{equation}
d_{\Delta}\left(  x,y\right)  <\delta_{\hat{\varepsilon}}\implies\left\vert
u_{t}\left(  a,x\right)  -u_{t}\left(  a,y\right)  \right\vert <\hat
{\varepsilon}\quad\forall t\in T,\forall a\in A \label{eq:equ-u}%
\end{equation}
Consider the profile $m=\left(  m_{t}\right)  _{t\in T}$ of
message\ functions\ such that each $m_{t}:A\times\Delta\rightarrow\Delta$ is
defined to be such that%
\[
m_{t}\left(  a,x\right)  =x\qquad\forall a\in A,\forall x\in\Delta
\]
Note that in this case $\left(  M,d\right)  =\left(  \Delta,d_{\Delta}\right)
$. Clearly, $m=\left(  m_{t}\right)  _{t\in T}$\ is a family of functions
which is equicontinuous with respect to the second argument. Given
$\varepsilon>0$, consider\ $\delta_{\varepsilon/2}>0$ as in (\ref{eq:equ-u}).
By Corollary \ref{cor:eps-sce}, we have that there exists a self-confirming
$\delta_{\varepsilon/2}/2$-equilibrium $\sigma\in\Sigma$, that is, there
exists $\beta\in\Delta^{T}$ such that%
\begin{align*}
&  1=\lambda\left(  \left\{  t\in T\ \left\vert \
\begin{array}
[c]{l}%
u_{t}\left(  \sigma\left(  t\right)  ,\beta\left(  t\right)  \right)  \geq
u_{t}\left(  a,\beta\left(  t\right)  \right)  \quad\forall a\in A\\
d\left(  m_{t}\left(  \sigma\left(  t\right)  ,\beta\left(  t\right)  \right)
,m_{t}\left(  \sigma\left(  t\right)  ,\lambda_{\sigma}\right)  \right)
\leq\delta_{\varepsilon/2}/2
\end{array}
\right.  \right\}  \right) \\
&  =\lambda\left(  \left\{  t\in T\ \left\vert \
\begin{array}
[c]{l}%
u_{t}\left(  \sigma\left(  t\right)  ,\beta\left(  t\right)  \right)  \geq
u_{t}\left(  a,\beta\left(  t\right)  \right)  \quad\forall a\in A\\
d_{\Delta}\left(  \beta\left(  t\right)  ,\lambda_{\sigma}\right)  \leq
\delta_{\varepsilon/2}/2
\end{array}
\right.  \right\}  \right)
\end{align*}
Define by $O$ the set of \textquotedblleft optimizing\textquotedblright%
\ players%
\[
O=\left\{  t\in T\ \left\vert \
\begin{array}
[c]{l}%
u_{t}\left(  \sigma\left(  t\right)  ,\beta\left(  t\right)  \right)  \geq
u_{t}\left(  a,\beta\left(  t\right)  \right)  \quad\forall a\in A\\
d_{\Delta}\left(  \beta\left(  t\right)  ,\lambda_{\sigma}\right)  \leq
\delta_{\varepsilon/2}/2
\end{array}
\right.  \right\}
\]
Since $u$ satisfies (\ref{eq:equ-u}), note that if $t\in O$, then\ we have
that $d_{\Delta}\left(  \beta\left(  t\right)  ,\lambda_{\sigma}\right)
\leq\delta_{\varepsilon/2}/2<\delta_{\varepsilon/2}$ which implies that for
each $a\in A$%
\[
\left\vert u_{t}\left(  \sigma\left(  t\right)  ,\beta\left(  t\right)
\right)  -u_{t}\left(  \sigma\left(  t\right)  ,\lambda_{\sigma}\right)
\right\vert <\frac{\varepsilon}{2}\text{ and }\left\vert u_{t}\left(
a,\beta\left(  t\right)  \right)  -u_{t}\left(  a,\lambda_{\sigma}\right)
\right\vert <\frac{\varepsilon}{2}%
\]
Since $t\in O$, we can conclude that%
\begin{align*}
u_{t}\left(  \sigma\left(  t\right)  ,\lambda_{\sigma}\right)   &
>u_{t}\left(  \sigma\left(  t\right)  ,\beta\left(  t\right)  \right)
-\frac{\varepsilon}{2}\geq u_{t}\left(  a,\beta\left(  t\right)  \right)
-\frac{\varepsilon}{2}\\
&  >u_{t}\left(  a,\lambda_{\sigma}\right)  -\frac{\varepsilon}{2}%
-\frac{\varepsilon}{2}=u_{t}\left(  a,\lambda_{\sigma}\right)  -\varepsilon
\quad\forall a\in A
\end{align*}
Since $t$ was arbitrarily chosen in $O$, we have that%
\[
O\subseteq\left\{  t\in T\ \left\vert \ u_{t}\left(  \sigma\left(  t\right)
,\lambda_{\sigma}\right)  \geq u_{t}\left(  a,\lambda_{\sigma}\right)
-\varepsilon\quad\forall a\in A\right.  \right\}
\]
Since $O$ has mass $1$, it follows that $\sigma\in\Sigma$ is an $\varepsilon
$-Nash equilibrium.\hfill$\blacksquare$

\bigskip

\noindent\textbf{Proof of Corollary\ \ref{cor:eps-pce}}. It is enough to
observe that a nonatomic game with \textit{a neighborhood structure }can be
mapped into a nonatomic game with \textit{estimation}\ feedback where $f$ is
defined as in (\ref{eq:fed-nei-str}).\ With this identification, an
$\varepsilon$-estimated equilibrium is a peer-confirming $\varepsilon
$-equilibrium. By Theorem \ref{thm:mai}, it is then enough to show that
$f=\left(  f_{t}\right)  _{t\in T}$\ is a family of functions which is
equicontinuous with respect to the third argument. But, note that%
\[
\left\vert f_{t}\left(  a,\gamma,x\right)  -f_{t}\left(  a,\gamma,y\right)
\right\vert =\left\vert d_{\Delta}\left(  \gamma,x\right)  -d_{\Delta}\left(
\gamma,y\right)  \right\vert \leq d_{\Delta}\left(  x,y\right)  \quad\forall
t\in T,\forall a\in A,\forall\gamma\in\Delta
\]
trivially proving equicontinuity with respect to the third argument of
$f$.\hfill$\blacksquare$

\bigskip

We conclude by proving Corollary \ref{cor:eps-bne}. But, before doing so, we
need to make an intermediate observation. Consider a statistical divergence
$D$. Recall that $D:\Delta\times\Delta^{o}\rightarrow\left[  0,\infty\right)
$ is a jointly convex and continuous function. Denote by $\mathcal{K}$ the
collection of all nonempty compact sets of $\Delta$.\ We endow $\mathcal{K}$
with the Hausdorff distance (see, e.g., Aliprantis and Border \cite[Chapter 3,
Sections 16 and 17]{AB}).\ We denote by\ $\mathcal{\bar{Q}}$\ a compact set of
$\mathcal{K}$ such that each $Q\in\mathcal{\bar{Q}}$ is a nonempty, convex,
and compact subset of $\Delta^{o}$. Given $x\in\Delta$ and $Q\in
\mathcal{\bar{Q}}$, consider the minimization problem%
\[
\min D\left(  x||y\right)  \text{ sub to }y\in Q
\]
Define $\mu:\Delta\times\mathcal{\bar{Q}}\rightrightarrows\Delta\mathcal{\ }%
$to be the solution correspondence of this\ minimization problem, that is, for
each $x\in\Delta$ and for each $Q\in\mathcal{\bar{Q}}$,%
\[
\mu\left(  x,Q\right)  =\left\{  z\in\Delta:z\in Q\text{ and }D\left(
x||z\right)  =\min_{y\in Q}D\left(  x||y\right)  \right\}
\]
By Berge's maximum theorem, note that $\mu$ is upper hemicontinuous when
$\Delta\times\mathcal{\bar{Q}}$ is endowed with the product topology. In
particular, if $D$ is strictly convex with respect to the second argument,
$\mu$ is single-valued, that is, $\mu$ is a continuous function. Finally,
define the map $g:\Delta\times\Delta\times\mathcal{\bar{Q}}\rightarrow\left[
0,\infty\right)  $ by%
\[
g\left(  \beta,x,Q\right)  =d_{\Delta}\left(  \beta,\mu\left(  x,Q\right)
\right)  \qquad\forall\beta,x\in\Delta,\forall Q\in\mathcal{\bar{Q}}%
\]
Since $\mu$ is a continuous function, it follows that $g$ is continuous when
$\Delta\times\Delta\times\mathcal{\bar{Q}}$ is endowed with the product
topology. By Aliprantis and Border \cite[Corollary 3.31]{AB}\ and since
$\Delta\times\Delta\times\mathcal{\bar{Q}}$ is a compact metric space, $g$ is
uniformly continuous.

\bigskip

\noindent\textbf{Proof of Corollary\ \ref{cor:eps-bne}}. Set $\mathcal{\bar
{Q}}=\operatorname*{cl}\mathcal{Q}$. By point 4\ of Remark \ref{rmk:eps-BNE},
note that $\mathcal{\bar{Q}}$\ is a compact subset of $\mathcal{K}$ such that
each $Q\in\mathcal{\bar{Q}}$ is a nonempty, convex, and compact subset of
$\Delta^{o}$. For each $t\in T$ define $f_{t}:A\times\Delta\times
\Delta\rightarrow\left[  0,\infty\right)  $ by%
\begin{equation}
f_{t}\left(  a,\gamma,x\right)  =g\left(  \gamma,x,Q_{t}\right)  \qquad\forall
a\in A,\forall\gamma,x\in\Delta\label{eq:mod-sta}%
\end{equation}
It is then enough to observe that a nonatomic game with \textit{model
misspecification }can be mapped into a nonatomic game with \textit{estimation}%
\ feedback where $f$ is defined as in (\ref{eq:mod-sta}) and $\Pi=\left\{
T_{1}\right\}  $.\footnote{Thus, $m=1$, $T_{1}=T$, and $\pi\left(  t\right)
=1$ for all $t\in T$.}\ With this identification, an $\varepsilon$-estimated
equilibrium is an $\varepsilon$-BNE. By Theorem \ref{thm:mai}, it is then
enough to show that $f=\left(  f_{t}\right)  _{t\in T}$\ is a family of
functions which is equicontinuous with respect to the third argument. Since
$g$ is uniformly continuous, the statement is trivially true.\hfill
$\blacksquare$

\bigskip

The proof of the last two points of Remark \ref{rmk:eps-BNE}\ is routine.
Thus, we conclude by only proving point 2.

\bigskip

\noindent\textbf{Proof of point 2 of Remark \ref{rmk:eps-BNE}}. Set
$\mathcal{\bar{Q}}=\operatorname*{cl}\mathcal{Q}$. Note that $\mathcal{\bar
{Q}}$\ is a compact subset of $\mathcal{K}$ such that each $Q\in
\mathcal{\bar{Q}}$ is a nonempty, convex, and compact subset of $\Delta^{o}$.
For each $t\in T$ define $f_{t}:A\times\Delta\times\Delta\rightarrow\left[
0,\infty\right)  $ as in the proof of Corollary \ref{cor:eps-bne}, that is,%
\[
f_{t}\left(  a,\gamma,x\right)  =g\left(  \gamma,x,Q_{t}\right)  \qquad\forall
t\in T,\forall a\in A,\forall\gamma,x\in\Delta
\]
Since $g$ is continuous and $\Delta\times\Delta\times\mathcal{\bar{Q}}$ is
compact, observe that $g\geq0$ takes a maximum value $M\geq0$. Define the
profile of feedback functions $\tilde{f}$ to be such that for each $t\in T$%
\[
\tilde{f}_{t}\left(  a,\gamma,x\right)  =\left\{
\begin{array}
[c]{cc}%
f_{t}\left(  a,\gamma,x\right)  & \gamma\in Q_{t}\\
M+1 & \gamma\not \in Q_{t}%
\end{array}
\right.  \qquad\forall a\in A,\forall\gamma,x\in\Delta
\]
Note that each $\tilde{f}_{t}$ satisfies (\ref{eq:wea-gro}). By the proof of
Corollary \ref{cor:eps-bne}, $f=\left(  f_{t}\right)  _{t\in T}$ is a family
of functions which is equicontinuous with respect to the third argument. It
follows that for each $\varepsilon>0$ there exists $\delta_{\varepsilon}>0$
such that%
\[
d_{\Delta}\left(  x,y\right)  <\delta_{\varepsilon}\implies\left\vert
f_{t}\left(  a,\gamma,x\right)  -f_{t}\left(  a,\gamma,y\right)  \right\vert
<\varepsilon\quad\forall t\in T,\forall a\in A,\forall\gamma\in\Delta
\]
Consider $x,y\in\Delta$ such that $d_{\Delta}\left(  x,y\right)
<\delta_{\varepsilon}$ and consider $t\in T$, $a\in A$, and $\gamma\in\Delta$.
We have two cases,\ either $\gamma\in Q_{t}$ or $\gamma\not \in Q_{t}$. In the
first case, $\left\vert \tilde{f}_{t}\left(  a,\gamma,x\right)  -\tilde{f}%
_{t}\left(  a,\gamma,y\right)  \right\vert =\left\vert f_{t}\left(
a,\gamma,x\right)  -f_{t}\left(  a,\gamma,y\right)  \right\vert <\varepsilon$,
while in the second case $\left\vert \tilde{f}_{t}\left(  a,\gamma,x\right)
-\tilde{f}_{t}\left(  a,\gamma,y\right)  \right\vert =\left\vert M+1-\left(
M+1\right)  \right\vert =0<\varepsilon$. Since $t$, $a$, and $\gamma$ were
chosen arbitrarily, it follows that\ $\tilde{f}=\left(  \tilde{f}_{t}\right)
_{t\in T}$ is a family of functions which is equicontinuous with respect to
the third argument. Next, we can consider the nonatomic game with estimation
feedback\ $\left(  \left(  T,\lambda\right)  ,A,u,\left(  \Pi,\pi\right)
,\tilde{f}\right)  $\textit{, }where $\Pi=\left\{  T_{1}\right\}
$.\footnote{Thus, $m=1$, $T_{1}=T$, and $\pi\left(  t\right)  =1$ for all
$t\in T$.}\ By Theorem \ref{thm:mai}, we have that for each $\tilde
{\varepsilon}>0$\ there exists an $\tilde{\varepsilon}$-estimated equilibrium
$\sigma$ for this game, that is, there exists $\beta\in\Delta^{T}$\ such that%
\[
\lambda\left(  \left\{  t\in T\ \left\vert \
\begin{array}
[c]{l}%
u_{t}\left(  \sigma\left(  t\right)  ,\beta\left(  t\right)  \right)  \geq
u_{t}\left(  a,\beta\left(  t\right)  \right)  \quad\forall a\in A\\
\tilde{f}_{t}\left(  \sigma\left(  t\right)  ,\beta\left(  t\right)
,\lambda_{\sigma}\right)  \leq\tilde{\varepsilon}%
\end{array}
\right.  \right\}  \right)  =1
\]
If given $\varepsilon>0$\ we define $\tilde{\varepsilon}=\frac{\min\left\{
M+1,\varepsilon\right\}  }{2}>0$, since $\tilde{\varepsilon}<M+1,\varepsilon$,
then we have that%
\begin{align*}
&  \left\{  t\in T\ \left\vert \
\begin{array}
[c]{l}%
u_{t}\left(  \sigma\left(  t\right)  ,\beta\left(  t\right)  \right)  \geq
u_{t}\left(  a,\beta\left(  t\right)  \right)  \quad\forall a\in A\\
\tilde{f}_{t}\left(  \sigma\left(  t\right)  ,\beta\left(  t\right)
,\lambda_{\sigma}\right)  \leq\tilde{\varepsilon}%
\end{array}
\right.  \right\} \\
&  \subseteq\left\{  t\in T\ \left\vert \
\begin{array}
[c]{l}%
u_{t}\left(  \sigma\left(  t\right)  ,\beta\left(  t\right)  \right)  \geq
u_{t}\left(  a,\beta\left(  t\right)  \right)  \quad\forall a\in A\\
f_{t}\left(  \sigma\left(  t\right)  ,\beta\left(  t\right)  ,\lambda_{\sigma
}\right)  \leq\tilde{\varepsilon}\text{ and }\beta\left(  t\right)  \in Q_{t}%
\end{array}
\right.  \right\} \\
&  \subseteq\left\{  t\in T\ \left\vert \
\begin{array}
[c]{l}%
u_{t}\left(  \sigma\left(  t\right)  ,\beta\left(  t\right)  \right)  \geq
u_{t}\left(  a,\beta\left(  t\right)  \right)  \quad\forall a\in A\\
d_{\Delta}\left(  \beta\left(  t\right)  ,\operatorname{argmin}_{z\in Q_{t}%
}D\left(  \lambda_{\sigma}||z\right)  \right)  \leq\tilde{\varepsilon}\text{
and }\beta\left(  t\right)  \in Q_{t}%
\end{array}
\right.  \right\} \\
&  \subseteq\left\{  t\in T\ \left\vert \
\begin{array}
[c]{l}%
u_{t}\left(  \sigma\left(  t\right)  ,\beta\left(  t\right)  \right)  \geq
u_{t}\left(  a,\beta\left(  t\right)  \right)  \quad\forall a\in A\\
d_{\Delta}\left(  \beta\left(  t\right)  ,\operatorname{argmin}_{z\in Q_{t}%
}D\left(  \lambda_{\sigma}||z\right)  \right)  \leq\varepsilon\text{ and
}\beta\left(  t\right)  \in Q_{t}%
\end{array}
\right.  \right\}
\end{align*}
yielding the statement.\hfill$\blacksquare$

\end{document}